\documentclass[final,5p,times,twocolumn,authoryear]{elsarticle}

\usepackage{amssymb}
\usepackage{lipsum}
\usepackage{graphicx}	% Including figure files
\usepackage{amsmath}	% Advanced maths commands
\usepackage{longtable, tabu}  % Print long tables
\usepackage{url}
\usepackage{caption}
\usepackage{hyperref}

\hypersetup{
    colorlinks=true,
    linkcolor=blue,
    filecolor=magenta,      
    urlcolor=cyan,
}

\usepackage{xcolor, ulem}
\newcommand\hl{\bgroup\markoverwith{\textcolor{yellow}{\rule[-.5ex]{2pt}{2.5ex}}}\ULon}

\newcommand{\msun}{{\rm M}_{\odot}}
\usepackage{mathrsfs} %  to use Raph Smith’s Formal Script font in mathematics \mathscr{}, to replace \mathcal{}

\usepackage{xspace}

\newcommand{\feh}{\rm{[Fe/H]}\xspace}

\journal{New Astronomy}

\begin{document}

\begin{frontmatter}

\title{A Multiband Photometric Study of RR Lyrae Stars in M53 (NGC 5024)}

\author[inst1]{Shantanu A. Gaur}

\author[inst3]{Nitesh Kumar\corref{cor1}}
\ead{niteshchandra039@gmail.com}

\author[inst2]{Anupam Bhardwaj}

\author[inst1]{Aasheesh Raturi}

\cortext[cor1]{Corresponding author}

\affiliation[inst1]{organization={Dolphin (PG) Institute of Biomedical and Natural Sciences}, 
            city={Dehradun},
            postcode={248007}, 
            state={Uttarakhand},
            country={India}}
\affiliation[inst3]{organization={Department of Physics, Applied Science Cluster, University of Petroleum and Energy Studies (UPES)}, 
            city={Dehradun},
            postcode={248007}, 
            state={Uttarakhand},
            country={India}}

\affiliation[inst2]{organization={Inter-University Centre for Astronomy and Astrophysics (IUCAA)}, 
            addressline={Post Bag 4, Ganeshkhind}, 
            city={Pune},
            postcode={411007}, 
            state={Maharashtra},
            country={India}}
            
% \affiliation[inst4]{organization={Department of Physics and Astrophysics, University of Delhi}, 
%             city={Delhi},
%             postcode={110007}, 
%             state={Delhi},
%             country={India}}

%%Research highlights
% \begin{highlights}
% \item A comprehensive (UBVRI) photometric study was performed on RR Lyrae stars in Messier 53 (NGC 5024) using archival data spanning over 22 years, which provided an excellent time baseline for accurate period determination.
% \item The refined periods of RR Lyrae stars and high RRc fraction ($54.7\%$) confirm M53 as an Oosterhoff II globular cluster, with RRab stars following the OoII locus in period-amplitude diagrams.
% \item Accurate pulsation properties were used to derive I band period-luminosity and multi-band period-Wesenheit relations, which were compared against theoretical predictions.
% \item A weighted mean distance modulus of $16.242 \pm 0.050$ mag was determined, yielding a cluster distance of $17.717 \pm 0.408$ kpc, consistent with recent Gaia-based estimates.
% \end{highlights}

\begin{abstract}
We present a multiband (UBVRI) time-series photometric study of RR Lyrae (RRL) stars in the globular cluster Messier 53 (NGC 5024) to refine their pulsation properties and determine a precise cluster distance. The archival photometric data includes images taken over 22 years and 3 months using different optical telescopes, providing an excellent time baseline to investigate light curves of variable stars. Using Lomb-Scargle periodogram, we derived accurate periods for 29 fundamental-mode (RRab) and 35 first-overtone (RRc) RRLs. Template-fitting to phase-folded light curves provided robust mean magnitudes and amplitudes. The refined periods confirm M53 as an Oosterhoff II cluster, with a mean period of 0.649 days for RRab and 0.346 days for RRc, and a high RRc fraction (54.7\%). Most RRLs align with the horizontal branch in the color-magnitude diagram, while a few outliers result from blending effects. Period-amplitude diagrams show RRab stars following the Oosterhoff II locus. We derived I-band period-luminosity and multi-band period-Wesenheit relations, comparing them with theoretical predictions. A weighted mean distance modulus of 16.242 $\pm$ 0.05 mag yields a cluster distance of 17.717 $\pm$ 0.408 kpc, in agreement with recent estimates based on parallaxes from Gaia data.
\end{abstract}

%%Graphical abstract
%\begin{graphicalabstract}
%\includegraphics{grabs}
%\end{graphicalabstract}

\end{frontmatter}

%\tableofcontents

%% \linenumbers

%% main text

\section{Introduction}
% RR Lyrae stars (RRL) are low mass ($\sim $ 0.55 - 0.8 $\msun$) stars \citep{cassisi_evolutionary_2021}, currently in their core-helium burning stage and are found at the intersection of instability strip and horizontal branch (HB) in Hertzsprung-Russell (HR) diagram. Their pulsation period ranges between 0.2 - 1.2 days \citep{stringer_identification_2019}. RRLs are old (age $\approx$ 10 Gyrs, \citealt{savino_a_age_2020}) metal-poor population-II type stars that make upto 80\% of the variables in a globular cluster \citep{xxx}. RRLs have a well defined period luminosity relation at the near infrared wavelengths that makes them useful for the distance calibration in nearby clusters. They are also used as tracer of old stellar population in their host cluster \citep{kunder_rr_2013} and can be used for the detailed study the stellar evolution \citep{catelan_horizontal_2009}. 

RR Lyrae stars (RRLs) are low-mass stars ($\sim 0.55-0.8 \msun$) currently in the core-helium burning phase, located at the intersection of the instability strip and horizontal branch (HB) in the Hertzsprung-Russell (HR) diagram \citep{cassisi_evolutionary_2021, kumar_extraction_2025}. Their pulsation periods range from 0.2 to 1.2 days \citep{stringer_identification_2019}. These stars are old ($\approx 10$ Gyrs, \citealt{sarajedini_rr_2006, savino_a_age_2020}), metal-poor, population II stars, and they account for up to 80\% of variable stars in globular clusters \citep{clement_variable_2001}. RRLs have a well-defined period-luminosity relation in the near-infrared, making them valuable for distance calibration in nearby clusters. They also serve as tracers of old stellar populations in their host clusters \citep{kunder_rr_2013} and provide insight into stellar evolution \citep{catelan_horizontal_2009}. RRLs exhibit very little dispersion in their visual magnitudes within the same cluster \citep{arellano_ferro_ccd_2017}, but their magnitudes vary across different clusters due to metallicity differences \citep{jones_template_1996, molnar_first_2021}.

Messier 53 (M53 or NGC 5024) is a globular cluster located in the constellation Coma Berenices, with coordinates RA $13^{\rm h}12^{\rm m}55.25^{\rm s}$ and Declination $+18^{\circ}10^{'}5.4^{''}$ \citep{goldsbury_acs_2010}. It lies at a distance of $18.0 \pm 0.4$ kpc from the Galactic center \citep{arellano_ferro_globular_2024}. Due to its sufficiently large longitude from the Galactic plane, it experiences low interstellar extinction with E(B-V) = 0.02 \citep{safonova_variables_2011, arellano_ferro_unusually_2012}. The estimated age of the M53 is $13.31^{+0.66}_{-0.57}$ Gyr \citep{valcin_inferring_2020} and the metallicity ($\feh$) from spectroscopic and photometric measurements are $-2.03 \pm 0.04$ and $-1.87 \pm 0.02$ respectively \citep{jurcsik_photometric_2023}.

M53 contains $109$ variable stars, of which $64$ are confirmed RR Lyrae stars. The other variables include SX Phoenicis, long-period variables, semi-regular variables, and other suspected variables \citep[][hereafter CC01]{clement_variable_2001}. Some of the variables in the record are also suspected to be misidentified constants. The variables of M53 have been the subject of various studies in the past. All these earlier studies have focused on either V and I bands \citep{kopacki_variable_2000, catelan_evolutionary_2004, arellano_ferro_exploring_2011, bramich_investigation_2012}, or the R-band \citep{safonova_variables_2011}. \cite{bhardwaj_rr_2021} investigated the RRLs of M53 in $JHK_s$ bands of the NIR wavelength region. 

% In this paper, we studied the pulsation properties of RR Lyrae variables of M53 in a wider range of wavelengths: U, B, V, R, and I bands of the Landolt scale \citep{landolt_broadband_1992}, which is the first such multi-band analysis of the RRLs in this cluster. We also revisited the already found periods of the variables and compared them with current findings, and extended the investigations to yet unexplored U and B bands. We established a Period Luminosity (PL) relationship in the I-band, and a Period-Wesenheit (PW) relationship in four Wesenheit bands and used them to calculate distances to the cluster M53.

In this paper, we investigated the pulsation properties of RR Lyrae variables in M53 across a broader range of wavelengths U, B, V, R, and I bands of the Landolt system \citep{landolt_broadband_1992}. This represents the first multi-band analysis of RRLs in this cluster. We revisited the previously reported periods of the variables, compared them with current results, and extended the study to the previously unexplored U and B bands. Furthermore, we established a Period–Luminosity (PL) relation in the I band and a Period–Wesenheit (PW) relation in four Wesenheit bands, which were then used to determine the distance to M53.

% This study begins with a description of the analysis of the photometric data in section \ref{sec:data}. In section \ref{sec:methods}, we discuss the process of period finding and template fitting, followed by the analysis of the light curves. In section \ref{sec:results}, we use the pulsation parameters obtained through template fitting to plot the period distribution, Bailey diagram, PL and PW relations, and calculate distances of the RRLs in M53. In section \ref{sec:discussion}, we discuss the obtained results and compare them with the literature. In the last section \ref{sec:conclusion}, we offer a brief summary to conclude our study.
This study begins with a description of the photometric data in Section~\ref{sec:data}. Section~\ref{sec:methods} describes the procedures for period determination and template fitting, followed by the light-curve analysis. In Section~\ref{sec:results}, we present the pulsation parameters derived from template fitting and use them to examine the period distribution, Bailey diagram, and the PL and PW relations, as well as to estimate the distance to M53. Section~\ref{sec:discussion} provides a comparison of results with those in the literature, and finally, Section~\ref{sec:conclusion} summarizes the main findings of this study.

\section{Data}\label{sec:data}
The photometric dataset was obtained from public archives that span a period of 22 years and 3 months, starting from February 1996 to May 2018. The data consists of 869 different images that have been grouped into 24 observation sets. Table \ref{tab:observation} gives the log of observations and a detail of the different photometric filters used for the run. 
% In the observations, the R-band data is very limited and was mostly collected during the \texttt{int0506} run.

The images were reduced using \texttt{DAOPHOT/ALLSTAR/ALLFRAME} suite of programs and transformed observations from non-standard filters to the standard Johnson-Krons-Cousins system. The images were calibrated to the Johnson \texttt{UBV} and Krons-Cousins \texttt{RI} photometric system \citep{stetson_homogeneous_2000, stetson_homogeneous_2005, stetson_optical_2014, braga_distance_2015, braga_rr_2016} which is very close to the photometric system of Landolt \citep{landolt_broadband_1992}. The data also contained observations in Sloan \texttt{ugri} and Stromgren \texttt{yb} filters, which were then transformed to the standard Landolt system. The Sloan `r' and `i' bands transform perfectly to that of Landolt R and I and can be treated as equivalent. Similarly, the Stromgren b and y transform to Landolt B and V, respectively, very well. The Sloan g band, on the other hand, lies between the Landolt B and V bands, and it has been converted to either B or V depending on whether the Sloan g filter used in the observatory was closer to the Landolt B or V. The discrepancy between Sloan u or Stromgren u with Landolt U is slightly more.

We do not use data converted from Sloan and Stromgren filters for template fitting to estimate pulsation parameters, but use them to calculate the pulsation periods. While Sloan and Stromgren-based readings are useful in finding period and mean magnitudes, their amplitudes can often be off, and hence they have not been considered during the template fitting process.

\begin{table*}
    \caption{Details of the observations used for NGC 5024.}
    \centering
	\label{tab:observation}
	\begin{tabular}{lcccccccc} 
\hline
Observer ID & Images & Start Date                      & End Date             & U  & B  & V  & R  & I   \\
            &        & {[}year-month-day-hour-min-sec] & \multicolumn{1}{l}{} &    &    &    &    &     \\ 
\hline
bond2       & 8      & 1996-03-12 11:04:16             & 1996-03-12 11:37:23  & 2  & 2  & 2  & 0  & 2   \\
lee3        & 60     & 2011-05-26 03:43:37             & 2011-05-27 03:56:35  & 0  & 24 & 36 & 0  & 0   \\
int1304     & 39     & 2013-04-12 01:44:15             & 2013-04-13 01:43:32  & 30 & 9  & 0  & 0  & 0   \\
int0605     & 60     & 2006-05-31 21:29:05             & 2006-06-01 00:07:37  & 24 & 0  & 36 & 0  & 0   \\
arg02       & 6      & 2002-05-09 22:13:35             & 2002-05-09 22:39:38  & 0  & 2  & 2  & 2  & 0   \\
int1204     & 57     & 2012-04-24 23:51:12             & 2012-04-25 01:25:49  & 12 & 15 & 15 & 0  & 15  \\
bond11      & 16     & 1998-03-23 07:51:10             & 1998-03-23 09:45:04  & 4  & 4  & 4  & 0  & 4   \\
hannah      & 16     & 2002-03-28 05:16:39             & 2002-03-28 06:16:16  & 6  & 7  & 3  & 0  & 0   \\
wfi6        & 48     & 2002-02-19 07:38:47             & 2002-02-19 08:19:40  & 0  & 16 & 16 & 0  & 16  \\
benetti     & 38     & 2000-04-26 23:19:40             & 2000-04-27 01:27:15  & 0  & 10 & 10 & 0  & 18  \\
cf0102      & 86     & 2001-02-18 13:03:12             & 2001-02-18 15:32:58  & 0  & 24 & 25 & 0  & 37  \\
dahl        & 2      & 2014-06-25 01:13:09             & 2014-06-25 01:13:35  & 0  & 0  & 1  & 1  & 0   \\
spm1802     & 14     & 2018-02-25 10:22:56             & 2018-02-25 11:11:11  & 2  & 3  & 3  & 3  & 3   \\
int1504     & 39     & 2015-04-26 00:09:30             & 2015-04-26 20:46:36  & 0  & 15 & 15 & 0  & 9   \\
Y1005       & 80     & 2010-05-03 03:09:04             & 2010-05-08 03:30:14  & 0  & 24 & 32 & 0  & 24  \\
alf03       & 5      & 2003-05-02 01:18:54             & 2003-05-02 01:49:26  & 1  & 1  & 1  & 1  & 1   \\
int1704     & 12     & 2017-04-06 02:16:56             & 2017-04-06 02:23:42  & 0  & 6  & 6  & 0  & 0   \\
int1202     & 60     & 2012-02-24 02:27:10             & 2012-02-25 04:19:55  & 60 & 0  & 0  & 0  & 0   \\
int         & 18     & 1998-06-23 22:42:40             & 1998-06-23 23:39:41  & 0  & 6  & 6  & 0  & 6   \\
int0506     & 116    & 2005-06-09 22:32:26             & 2005-06-18 21:53:08  & 0  & 41 & 3  & 29 & 43  \\
int1805     & 43     & 2018-05-20 21:08:21             & 2018-05-22 22:21:47  & 10 & 6  & 21 & 6  & 0   \\
int1802     & 12     & 2018-02-22 02:23:08             & 2018-02-22 02:36:14  & 6  & 6  & 0  & 0  & 0   \\
west2       & 16     & 2005-05-05 04:18:54             & 2005-05-07 04:05:22  & 0  & 6  & 0  & 4  & 6   \\
int1502     & 18     & 2015-02-27 05:23:42             & 2015-02-27 05:35:05  & 0  & 9  & 9  & 0  & 0   \\
\hline
\end{tabular}
\end{table*}

\subsection{Filtering of Data}\label{sec:data selection}
For accurate period determination and analysis of light curve parameters of variable stars, errors should be kept at a minimum. The photometric observations of a target source include errors in the measurement, and only those measurements that lie below a certain threshold should be used so that variability of small amplitudes is also detected and not confused with dispersion due to inherent error. We adopted a limit of 0.07 mag as the maximum allowed error in all bands. This low threshold for error made the calculation of periods and pulsation properties more accurate.

We find that all of the variables have fewer observation data points in the R-band, while having a minimum of $35$ observations in all the other bands before filtration. The lack of data in the R-band led to difficulties in period calculation and template fitting. The highest number of readings in the R-band is 30, which occurs for 19 RRLs. We have only 3 readings in R-band for the variables V11, V12, V20, and V29. After filtering, the maximum number of R-band readings in any variable drops to 28. We note that on average, 81.45\% of data points are retained in the R-band. In the case of the U-band, 91.58\% of data points were of good quality. Similarly, in the B-band, an average of 90.39\% data had errors lower than the allowed limit. In the V-band, the average percentage of good points retained for the variables was 90.29\%. Lastly, for I-band, we retained 85.06\% of the data post-filtering. 
% To deal with the problem of a lack of observations in the R-band, we adopt the use of amplitude ratios to fit templates in the R-band by constraining their variation.

\section{Methods}\label{sec:methods}
We found periods for 64 RRLs using the Lomb-Scargle periodogram, followed by the determination of pulsation properties using template fitting. The sample includes 35 RRc and 29 RRab in the catalog by \cite{clement_variable_2001}, which was last updated in 2012 following the study by \cite{arellano_ferro_exploring_2011}. Before starting with the analysis, we filtered out the high error photometric data and proceeded with the calculation of amplitude ratios.

\subsection{Period Determination}\label{sec:period}
The photometric data of variables can show multiple periodicities, which arise due to cadence, along with the inherent periodicity of the source. Observation of targets is not continuous and has breaks, which are often regular. Cadence introduces artificial patterns in the data that can be confused as periodic frequencies by the period-determining algorithms \citep{saha_hybrid_2017}. A period finding algorithms, for e.g. \cite{graham_comparison_2013}, identify the strongest frequency present in the photometry of the source, which corresponds to the fundamental period of vibration.

We used the Lomb-Scargle (LS) periodogram \citep{lomb_least-squares_1976, scargle_studies_1982}, modified by \cite{vanderplas_periodograms_2015} to expand its applications to multiband data, for calculating the periods of RRLs by simultaneously using data in UBVRI bands. In the LS method, a fourier transform is performed on the data, which results in a power spectrum for given test frequencies. In the modified version, there is also a common base model built for all the bands, which treats the data as parts of a fourier series, assuming a common phase and period, and provides a periodogram based on the residuals forming in different bands compared to the base model. This results in a common high power frequency of variability for all the bands combined, corresponding to the fundamental period.

We utilized the \texttt{LombScargleMultiband} function from the \texttt{Gatspy} package in \texttt{AstroML}\footnote{\url{https://www.astroml.org/gatspy/periodic/lomb_scargle_multiband.html}.} within Python to perform multiband Lomb-Scargle period estimation. This function takes as input the time (in HJD), corresponding measured magnitudes, associated uncertainties, and filter information. The model was configured with a lower period bound of 0.002 days and an upper bound of 2 days. The estimated periods were then used to phase-fold the raw light curves. The phase ($\phi$) for any given time ($t$) relative to the initial epoch $t_0$ (defined as the epoch of maximum brightness) is computed using the equation \citep{kumar_predicting_2023}:  

\begin{equation}\label{eq:phase}
    \phi = \left[\frac{t-t_0}{\rm P}\right] - \texttt{Int}\left[\frac{t-t_0}{\rm P}\right].
\end{equation}  

\noindent Here, $P$ represents the derived period of the time-series data.

% \subsubsection{Period Determination for Problematic Variables}\label{sec:period_problematic}
% We compared the periods derived using Multiband-LS (MBLS) with the periods derived in literature \citep{arellano_ferro_exploring_2011, safonova_variables_2011, bhardwaj_rr_2021} and found that for 19 of RRL variables has at least 0.01 days difference. In such cases, we performed a segmenting approach to determine the correct period. The periods were re-determined for variables V2, V4, V12, V14, V17, V19, V20, V30, V32, V36, V40, V44, V53, V55, V57, V61, V62, V72, and V91. In all these cases there were more than 0.01 days of difference between the calculated period and the period provided in \cite{clement_variable_2001}. In Section \ref{sec:colormagnitude} we discovered that V64, V53, V61, V63, and V62 suffer with blending in light curves due high density of the region leading to inaccuracies. We segmented the observations for each of these variables and calculated the period for every segment. This segmenting approach led to the identification of any uncertainties in data, and the segment for which period was the most accurate was expanded to judge the duration for which data led to reliable period determination. We adopted the period closest to previous records and determined the largest possible segment from the photometric observations. 

\subsubsection{Period Determination for Problematic Variables}\label{sec:period_problematic}  

We compared the periods derived using the Multiband Lomb-Scargle (MBLS) method with those reported in the literature \citep{arellano_ferro_exploring_2011, safonova_variables_2011, bhardwaj_rr_2021} and found discrepancies of at least 0.01 days for 19 RR Lyrae variables. To address these inconsistencies, we employed a segmentation-based approach to refine the period estimates. The periods were re-evaluated for variables V2, V4, V12, V14, V17, V19, V20, V30, V32, V36, V40, V44, V53, V55, V57, V61, V62, V72, and V91, all of which exhibited deviations exceeding 0.01 days compared to the values reported in \cite{clement_variable_2001}.  

As described in Section \ref{sec:colormagnitude}, we identified that V64, V53, V61, V63, and V62 were affected by blending due to the high stellar density in their respective regions, leading to inaccuracies in their light curves. To mitigate these issues, we segmented the observational data for each of these variables and determined the period for each segment individually. This approach allowed us to assess data inconsistencies and identify the segment with the most reliable period estimate. The segment yielding the most accurate period was then extended to determine the duration over which the period remained stable. Ultimately, we adopted the period closest to previous records and selected the longest viable segment from the photometric observations to ensure reliability in the final period determination.

We found that segments excluding observations prior to HJD 2453000 and after HJD 2458000 resulted in significantly improved period determinations. The complete dataset spans HJD 2451000 to HJD 2459000, and the discrepancies in period estimation may be attributed to a lower number of observations or irregular cadence.  

\begin{figure*}
    \centering
    \includegraphics[width=\linewidth]{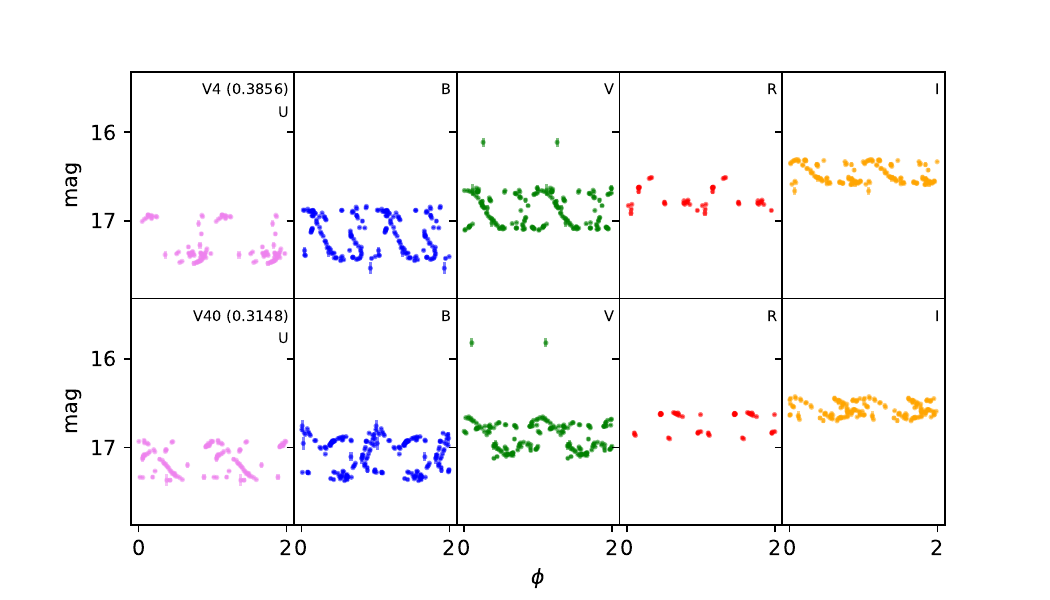}
    % \caption{Light curves for V4 (top row) and V40 (bottom row) RR Lyrae stars (RRLs) for which period analysis was performed by segmenting the observations. The columns, from left to right, show light curves in the U, B, V, R, and I bands, respectively. The calculated period reveals a typical RRc-type variation superimposed on noisy background data. For a clearer representation with template fits for both variables, refer to Figures \ref{img:temp1to10} and \ref{img:tempv32to42}.}
    \caption{Light curves of RR Lyrae stars V4 (top) and V40 (bottom) in the $U$, $B$, $V$, $R$, and $I$ bands. The derived periods show typical RRc-type variations with noisy background; see Figures~\ref{img:temp1to10} and \ref{img:tempv32to42} for template fits.}
    \label{img:phasesegment}
\end{figure*}

Tables \ref{tab:segment1} and \ref{tab:segment2} present representative examples of the segmented period calculation process for the variable stars V4 and V40. Table \ref{tab:correctperiods} summarizes the initially derived periods, the final selected periods after segmentation, and the corresponding periods reported in CC01 for the variables discussed in this section. Figure \ref{img:phasesegment} illustrates the phase-folded light curves for representative cases using the refined periods. The light curves exhibit the expected sinusoidal variations but also reveal the impact of noisy data corresponding to problematic observations.  

\begin{table}  
\centering  
% \caption{Period determination by segmenting the total observation time into distinct ranges for the variable star V4 (RRc; period in CC01: 0.3856 days). The period for each observational time range is computed, and the ranges are adjusted to identify the time span that provides the most accurate period estimation. The start and end dates are given in Heliocentric Julian Date (HJD).}  
\caption{Period determination for the variable star V4 (RRc; period in CC01: 0.3856 days) by dividing the total observation time into distinct ranges. For each range, the period is computed, and the ranges are adjusted to identify the interval yielding the most accurate estimate. Start and end dates are reported in Heliocentric Julian Date (HJD).}

\begin{tabular}{ccc}  
\hline  
Start HJD & End HJD     &  Calculated Period (days) \\  
\hline  
2450000 & 2459000 & 0.63897 \\  
2454000 & 2456000 & 0.38558 \\  
2456000 & 2458000 & 0.36373 \\  
2453000 & 2458000 & 0.38522 \\  
2456000 & 2459000 & 0.7337 \\  
2454000 & 2457000 & 0.38519 \\  
\hline  
\end{tabular}  
\label{tab:segment1}  
\end{table}

\begin{table}
    \centering
    \caption{Same as Table \ref{tab:segment1} but for RRc variable V40 (period in CC01: 0.3147).}    
    \begin{tabular}{ccc}
    \hline 
    Start HJD & End HJD     &  Calculated Period (days) \\
    \hline 
    2450000 & 2458000 & 0.46672 \\
    2450000 & 2454000 & 0.33301\\
    2454000 & 2456000 & 0.31430\\
    2456000 & 2459000 & 0.64856\\
    2453000 & 2456000 & 0.46105\\
    2456000 & 2458000 & 0.44603\\
    2454000 & 2457000 & 0.314823\\
    \hline
    \end{tabular}
    \label{tab:segment2}
\end{table}

\begin{table}
    \centering
    \caption{The updated periods of 19 variables described in Sec \ref{sec:period_problematic}. 
    % \noten{Need more significant digits. }
    }    
    \begin{tabular}{ccc}
    \hline Variable & Period (CC01) & Calculated Period (days)\\
    \hline
    V2 &  0.3862 & 0.38620\\
    V4 &  0.3856 & 0.38558\\
    V12 & 0.6126 & 0.61259\\
    V14 & 0.5453 & 0.54549\\
    V18 & 0.3361 & 0.33606\\
    V19 & 0.3910 & 0.39117\\
    V20 & 0.3842 & 0.38414\\
    V30 & 0.5355 & 0.53534\\
    V32 & 0.3904 & 0.39054\\
    V36 & 0.3732 & 0.37687\\
    V40 & 0.3147 & 0.31482\\
    V44 & 0.3749 & 0.37494\\
    V53 & 0.3891 & 0.33372\\
    V55 & 0.4433 & 0.44322\\
    V57 & 0.5683 & 0.53630\\
    V61 & 0.3795 & 0.30119\\
    V62 & 0.3745 & 0.35986\\
    V72 & 0.3407 & 0.38453\\
    V91 & 0.3024 & 0.30393\\
    \hline
    \end{tabular}
    \label{tab:correctperiods}
\end{table}

\subsection{Amplitude Ratio}\label{sec:ampratio}
We determined amplitude ratios with respect to V-band magnitudes and used them to constrain parameters during template fitting. Both RRab and RRc type variables exhibit constant amplitude ratios between one band and another with some dependence on metallicity \citep{inno_new_2015}. Amplitude ratios are helpful for amplitude determination in bands with poor observations provided the amplitude in some other known reference band \citep{jones_template_1996, soszynski_mean_2005, inno_new_2015, braga_rr_2016}. We calculated amplitude ratios for all bands with respect to the V-band amplitude since V-band, in almost all cases, was well sampled, particularly around the extrema. Following the same procedure as \cite{braga_rr_2016}, we used observations of only those bands for which the sampling was good and outliers were the least. 

% \begin{table}
%     \centering
%     \caption{Amplitude ratios calculated for RRab and RRc variables in M53.}
%     \begin{tabular}{ccc}
%          \hline Ratio & RRab & RRc \\
%          \hline $A_U/A_V$ & 1.11896 & 1.07874 \\
%           $A_B/A_V$ & 1.22133 & 1.15458 \\
%            $A_R/A_V$ & 0.67021 & 0.72506 \\
%             $A_I/A_V$ & 0.66839 & 0.68950 \\
%     \hline
%     \end{tabular}
%     \label{tab:ampratio}
% \end{table}

\begin{table*}
\centering
\caption{Mean Amplitude Ratios for RRab and RRc variables from different studies.}
\label{tab:ampratio}
\begin{tabular}{lcccccc}
\hline
\multicolumn{1}{c}{Ratio} & \multicolumn{2}{c}{\cite{kumar_multiwavelength_2024} (M3)} & \multicolumn{2}{c}{\cite{braga_rr_2016} ($\omega$ Cen)} & \multicolumn{2}{c}{This work (M53)} \\
\cline{2-7}
& RRab & RRc & RRab & RRc & RRab & RRc \\
\hline
$A_U/A_V$ & 1.05$\pm$0.21 & 1.06$\pm$0.16 & \multicolumn{2}{c}{...} & 1.11896 & 1.07874 \\
$A_B/A_V$ & 1.22$\pm$0.16 & 1.28$\pm$0.18 & 1.25$\pm$0.01 & 1.26$\pm$0.02 & 1.22133 & 1.15458 \\
$A_R/A_V$ & 0.72$\pm$0.35 & 1.00$\pm$0.42 & 0.80$\pm$0.03 & 0.77$\pm$0.02 & 0.67021 & 0.72506 \\
$A_I/A_V$ & 0.61$\pm$0.15 & 0.71$\pm$0.17 & 0.63$\pm$0.01 & 0.63$\pm$0.01 & 0.66839 & 0.68950 \\
\hline
\end{tabular}
\end{table*}

% Table \ref{tab:ampratio} shows the amplitude ratios $\frac{A_\lambda}{A_V}$ for both RRab and RRc, where $\lambda$ refers to the appropriate filter (U, B, R, or I). The ratios for RRab are higher than those of RRc in shorter wavelengths but seem nearly equal in longer wavelengths.

Table \ref{tab:ampratio} presents the amplitude ratios $\frac{A_\lambda}{A_V}$ for both RRab and RRc variables, where $\lambda$ corresponds to the respective filter (U, B, R, or I). The amplitude ratios for RRab are larger than those for RRc at shorter wavelengths, but they converge to nearly identical values at longer wavelengths.

\subsection{Template Fitting of RR Lyrae Light Curves} \label{sec:template fitting}
\subsubsection{Alignment of Light Curves} \label{sec:alignment}
We aligned the observed light curves (LCs) to a common epoch similar to the templates used to constrain variation in phase shifts during template fitting. Alignment of LCs in all the filters to a common epoch for a particular variable star helps in constraining the variation of phase shift between the template light curve and the variable light curve, which helps to avoid over-fitting or under-fitting of data, particularly in the presence of outliers.  

We chose the epoch of maximum brightness in the V-band as the reference point for the alignment. It is known that for RRab, there is a systematic shift in phase of maximum brightness towards higher phases as one moves towards the longer wavelengths \citep{bhardwaj_rr_2022}, but this effect is not considerable, and the differences in phase of maxima in any band from that of V-band is within the allowed levels of variation during template fit. V-band also enjoys the advantage that its data is well sampled for all variables, and only in a handful of cases there are outliers or gaps.  

We used g-band templates (refer to section \ref{sec:tempfit} for details of templates) to fit V-band light curves, and the best-fit template with the least chi-square spread was chosen as the reference for the alignment to a common epoch. We allowed a variation of free parameters, i.e. mean magnitude, amplitude, and phase shift, by 0.001 mags, 20\% mag, and 0.5, respectively to scale the templates as well as possible.

\subsubsection{Template fitting to aligned light curves} \label{sec:tempfit}

The method of template fitting (TF) has the advantage of allowing the calculation of pulsation parameters even in the presence of sparse data \citep{sesar_light_2009, gavrilchenko_mid-infrared_2014, hoffman_periods_2021}, which is problematic when using the method of Fourier Fitting (FF) to do the same \citep{kovacs_computation_2007}. While TF is more sensitive to outliers than FF, this problem can be solved by constraining the variation of free parameters in the fitting process and aligning the light curves to the templates (both having a maximum at the same phase), as was done in section \ref{sec:alignment}. TF becomes ineffective in the absence of a complete template set, extremely poor observations, or other methods as discussed by \cite{hoffman_periods_2021}.

We used the templates constructed using Stripe 82 SDSS data by \cite{sesar_light_2009} to fit the observed light curves using the chi-square minimization approach. The template set includes 11 templates in the u-band, 21 templates in the g-band, 20 templates in the r-band, 20 templates in the i-band, and 18 templates in the z-band for RRab. The number of templates in the same order for RRc was 1, 2, 2, 2, and 1, respectively. These templates were sequentially fitted to the observed light curve corresponding to the correct filter using the chi-square deviation ($\chi^2$, described in equation \ref{eq:chisq}) of the actual magnitude measurements ($m$) around the template fit magnitudes ($m_{\rm fit}$) to quantise the fits. 
\begin{equation}\label{eq:chisq}
    \chi^2 = \sum \frac{(m-m_{\rm fit})^2}{m_{\rm fit}}.
\end{equation}

While the use of non-Landolt filters is useful for period determination, we did not include them in the template fitting process since their amplitudes are unreliable due to errors introduced when they were converted. Filters like Sloan u, g, r, i and Stromgren b, y filters, do not accurately match the band passes of standard Landolt filters. Due to improper conversion, the errors can get transmitted to the mean magnitudes and amplitudes during template fitting. We observed improvements in fits when not using non-Landolt filters compared to when they were used.

\begin{figure}
    \centering
    \includegraphics[width=0.9\linewidth]{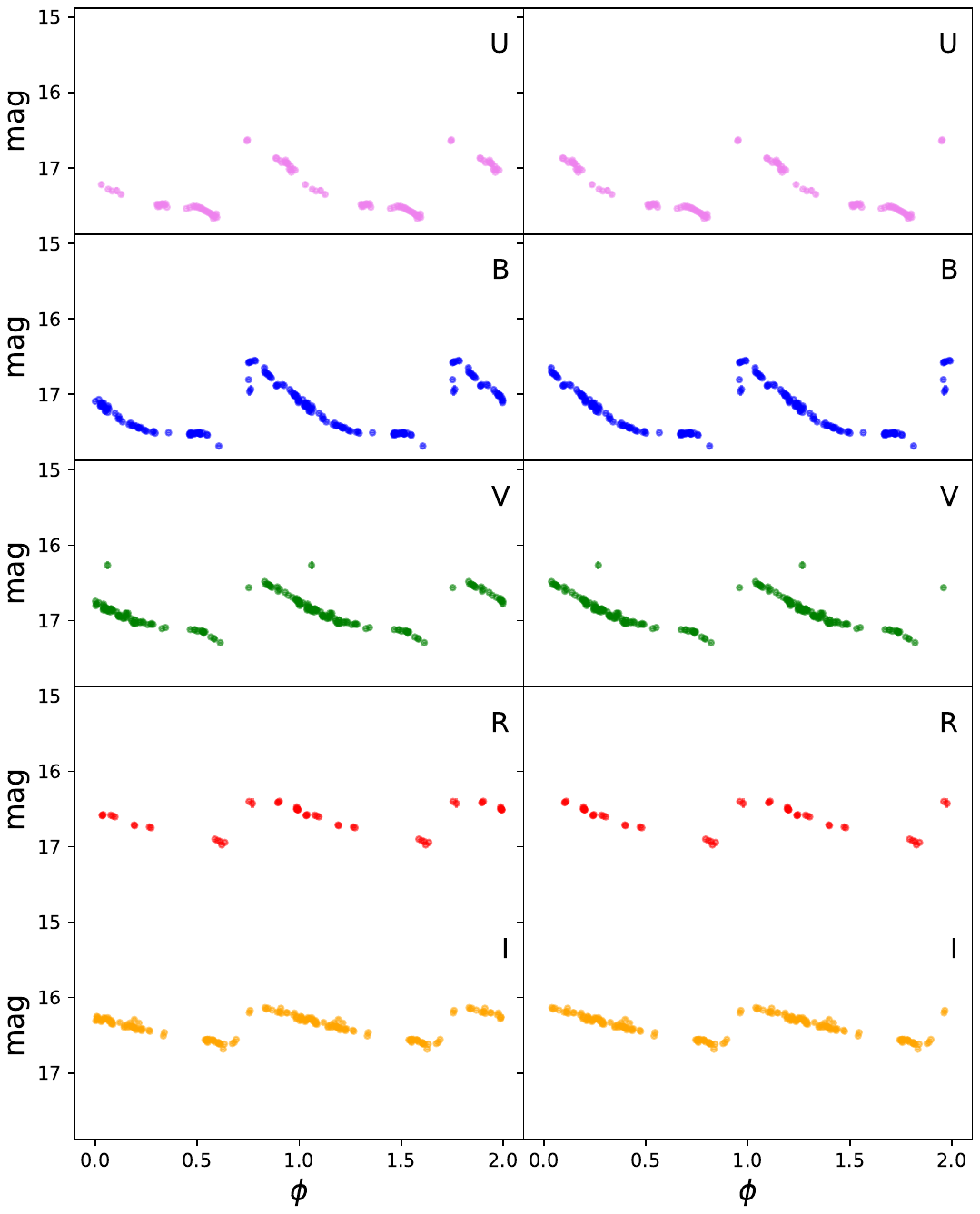}
    % \caption{The light curve for V4 (RRab) before and after aligning the maximum brightness in the V band to phase 0 (left and right panels, respectively).}
    \caption{Light curve of V4 (RRc) in the $V$ band, displayed before (left) and after (right) phasing the data such that maximum light occurs at phase~0.}
    \label{img:alignrr}
\end{figure}

We performed iterative TF for each variable in all bands keeping mean magnitude, amplitude, and phase shift as the free parameters. We allowed a variation of $\pm 0.1$ mag in mean magnitudes, of $\pm 30\%$ in amplitudes, and only $\pm 0.05$ in phase shifts for the first iteration. We reduced the variation in amplitude to $\pm 20\%$ for the successive iterations. At the end of each iteration, outliers outside the $3 \sigma$ level were removed. We performed a total of three iterations, which could be less if no outlier was detected.

We assumed a reference phase shift of zero since the LCs were practically aligned with the templates in terms of phases in the previous steps. We determined the reference mean magnitudes and amplitudes in the filters U, B, R, and I, for fitting as follows:

\begin{enumerate}
    \item We determined the amplitude by directly scaling the V-band amplitude through the amplitude ratio $\frac{A_{\lambda}}{A_V}$ for a filter $\lambda$ having either less than 20 observations or gaps around extrema. \cite{ngeow_zwicky_2022} had shown that template fits resulted in large residuals when the number of data points (N) was less than 10 and advised against applying template fitting to data with only 3 points. The same study showed that the average residual stabilized when more than 10 data points were available following the exponential decay function of the form, $f(n) = 0.187e^{-131N}$ ( where $N$ is the number of data points). The mean magnitude was calculated by taking the arithmetic mean.
    \item We determined amplitude by subtracting the minimum magnitude from the maximum magnitude in the filter when there was good phase coverage around the extrema. We determined the mean by averaging the maximum and minimum magnitudes to avoid clustering of data on one edge to influence the mean.
\end{enumerate}

We set a manual threshold for either the maxima, or the minima, or both, for RRc variables in the presence of outliers that could be problematic in amplitude determination. We did not exploit this step and applied it only to cases where the measurements clearly were erroneous, which we judged by the characteristic amplitude allowed for RRc variables (0.2 to 0.8 mag; \citealt{stringer_identification_2019}). Once all the available templates for a particular filter had been tried, the template with the smallest value of $\chi^2$ was taken to be the best template for the filter, and the free parameters of the final fit of that template were stored for the filter. 

We relied on the knowledge that SDSS bands are very close to the corresponding Landolt bands, and that the templates were normalised to fit the templates in different filters. Since the templates by \cite{sesar_light_2009} were based on ugriz SDSS bands, the g-band templates were fit to both Landolt B, and V bands of observed data due to the intersection of g-band with both the standard bands.

\subsubsection{Validation of template fits} \label{sec:validtemp}
We validated the template fits and their usefulness by comparing the light curve parameters derived from them with those derived from observed light curves. It has been observed that the V-band mean magnitudes ($V_{\rm mean}$) in RRLs do not vary much for a given cluster \citep{jameson_rr_1986}. This is true if the metallicity does not vary much within the cluster, which is true for most cases with the famous exception being $\omega$ Centauri \citep{braga_rr_2016}. We used the intensity averaged mean magnitudes ${\rm <}I{\rm >}$ (using \ref{eq:intavgmean}) instead of arithmetic mean magnitudes $I$ in any band. Many earlier studies support the use of the former mean magnitude for analysis, while some authors have also suggested the form $\frac{2}{3}{\rm <}I{\rm >} + \frac{1}{3}I$ for the mean magnitude \citep{cacciari_multicolor_2005}. 

\begin{equation} \label{eq:intavgmean}
    {\rm <}I{\rm >} = -2.5\log_{10}\left( \frac{\sum(10^{-0.4m})}{n(m)} \right).
\end{equation}
\noindent where m is the apparent magnitude of each star and n(m) are the number of observations. 

\begin{figure}
    \centering
    \includegraphics[width=\linewidth]{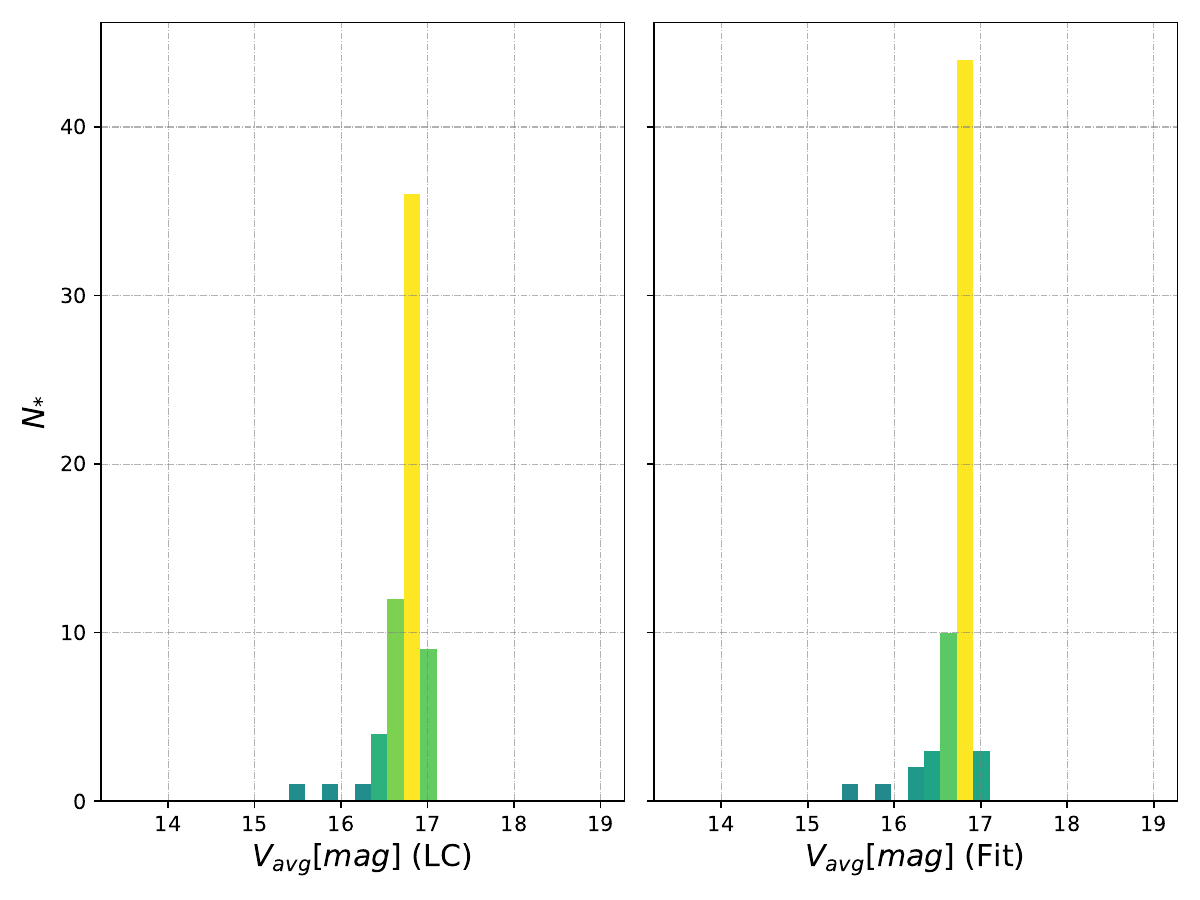}
    \caption{Intensity averaged $V_{mean}$ distribution for both light curve data (left panel) and template fitted data (right panel). It can be clearly seen that the scatter of $V_{mean}$ is much lower in template fit in agreement with the discussion of \ref{sec:validtemp}.}
    \label{img:vmeandistribution}
\end{figure}

We plotted a histogram showing the dispersion of intensity-averaged V-band means (${\rm <}V{\rm >}$) from raw LCs and also from templates as shown in figure \ref{img:vmeandistribution}. We clearly observe that the dispersion is very low for the case of template fits, compared to observed (raw) data.

\begin{figure}
    \centering
    \includegraphics[width=\linewidth]{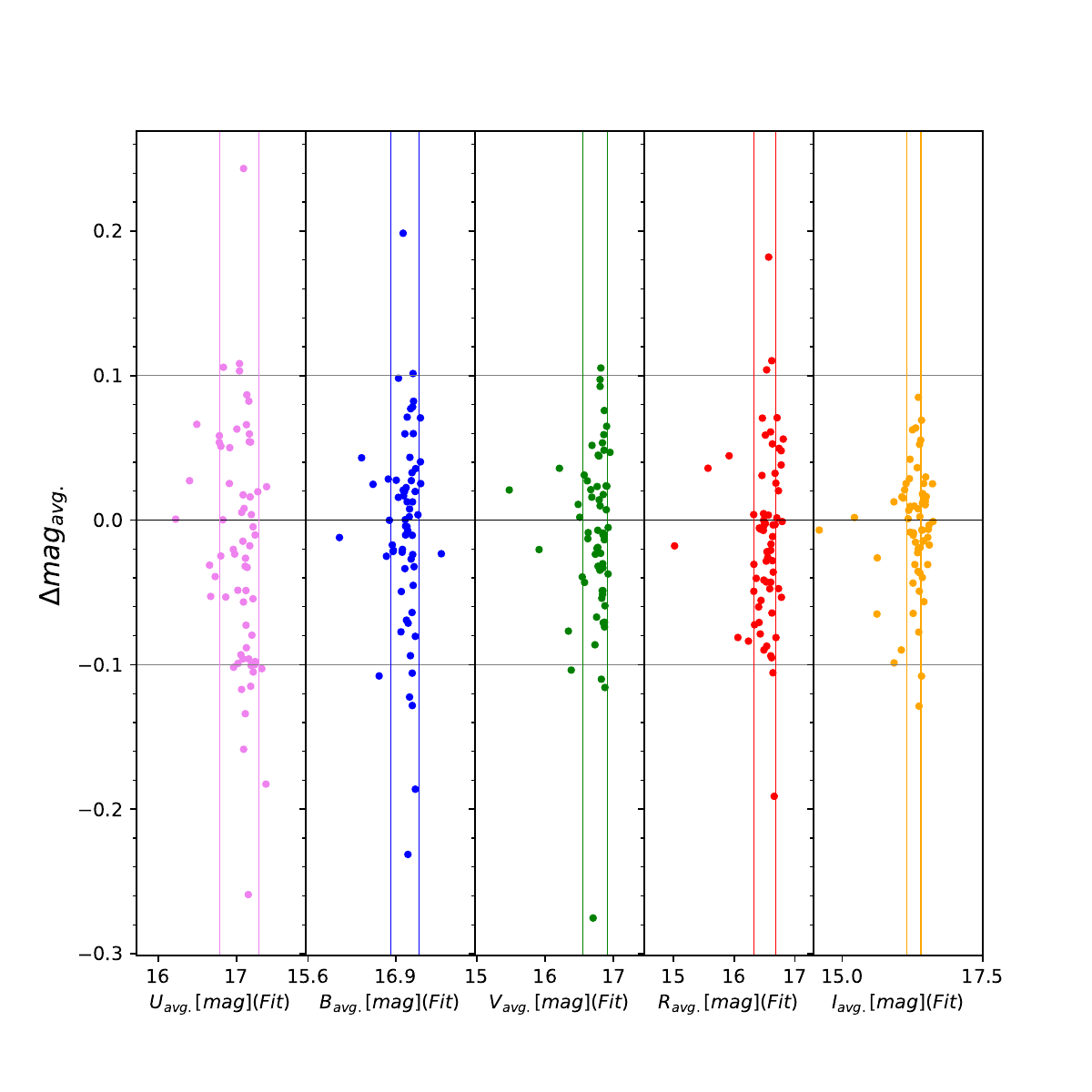}
    % \caption{Plot of difference in intensity averaged mean magnitudes $\Delta mag_{avg}$ from light curve data and template data against the intensity averaged mean magnitude in respective bands. The vertical colored lines in each panel represent the $3\sigma$ limits of the mean magnitude distribution. The panels from left to right show data for U, B, V, R, and I bands respectively.}
    \caption{Difference in intensity-averaged mean magnitudes ($\Delta mag_{\mathrm{avg}}$) between the observed light curve data and the template fits, plotted against the intensity-averaged mean magnitude in each band. The vertical colored lines mark the $3\sigma$ limits of the mean magnitude distribution. Panels from left to right correspond to the $U$, $B$, $V$, $R$, and $I$ bands.}

    \label{img:deltaintmean}
\end{figure}

We also plotted the deviation $\Delta V_{\rm mean} ( = V^r_{\rm mean} - V^t_{\rm mean})$ vs $V^t_{\rm mean}$ in the figure \ref{img:deltaintmean}. As can be seen from the figure most measurements in any band for template measured mean, cluster around a common value, with few outliers. While these deviations are small, the deviations are present due to the lower quality of observed data when compared to the complete phase coverage, and outlier free templates. The presence of the Blazhko effect is also one of the reasons for high scatters in light curves leading to considerable inaccuracies in predicting light curve parameters.

Table \ref{tab:parameter10} lists down the variables (column 1), calculated periods (column 2), type of RR Lyrae (column 3), intensity averaged mean magnitudes (columns 4, 8 and 12), their errors (column 5, 9, and 13), amplitudes (column 6, 10, and 14), and their errors (column 7, 11, and 15) for U, B and V bands respectively. Table \ref{tab:parameter11} refers to the same for R and I bands respectively.

\subsubsection{Flagging of templates}\label{sec-flags}
We assigned quality flags to the fitted templates, judging them by the mean RMS (equation \ref{eq:rms}) observed in the template with respect to the observed light curves. This was the final step in TF which assessed the quality of the template fits.  We observe that majority of the template fits are of high quality, and we identify light curves for which high RMS results due to reasons like Blazhko Effect, and blending in LCs (for instance in V17, V53, and others as discussed in section \ref{sec:colormagnitude}).
\begin{equation} \label{eq:rms}
    RMS = \sqrt{\frac{(m-m_{\rm fit})^2}{n(m)}}.
\end{equation}

\noindent where $m$ is the light curve magnitude, $m_{\rm fit}$ is the template magnitude, and $n (m)$ is the number of light curve magnitude readings.

\begin{figure}
    \centering
    \includegraphics[width=0.7\linewidth]{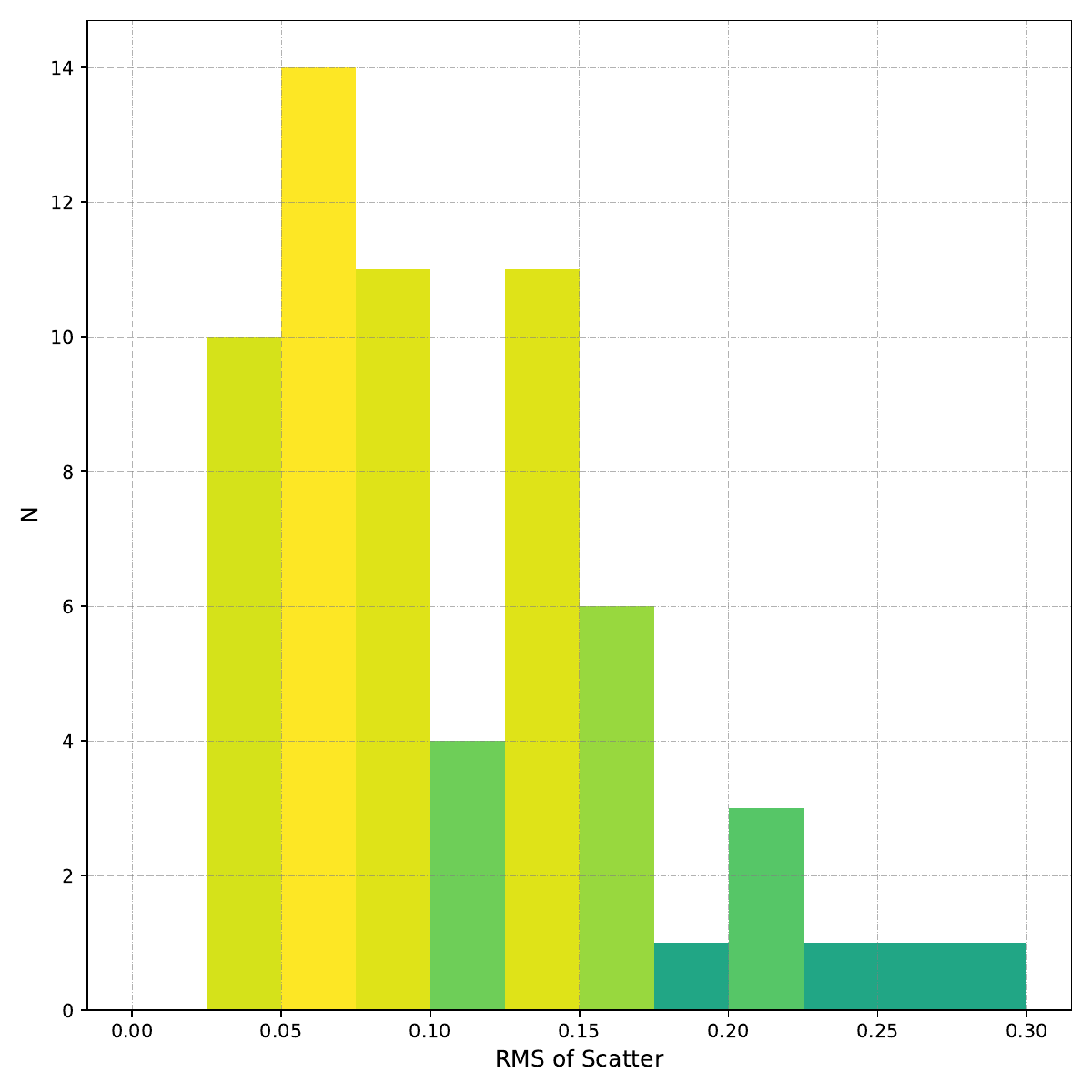}
    % \caption{Distribution of mean RMS scatter between light curve data and template fits for RRab and RRc variables of NGC 5024.}
    \caption{Distribution of mean RMS scatter between observed light curves and template fits for RRab and RRc variables in NGC 5024.}

    \label{img:rmsscatter}
\end{figure}

% A histogram showing the distribution of RMS values of the template around the observed light curve is plotted in figure \ref{img:rmsscatter}. We notice a majority of variables with less than 0.1 mean RMS. We defined three quality flags by visually observing the template fits along with the inferences drawn from the mean RMS histogram as follows:
% \begin{itemize}
%     \item \textbf{A}: Good LCs in all filters lead to nice template fits. 0 < RMS $\leq$ 0.1
%     \item \textbf{B}: Scatter in  LCs but clear periodicity and good template fit. 0.1 < RMS $\leq$ 0.2
%     \item \textbf{C}: Large scatter in LCs, leading to poor template fits. 0.2 < RMS $\leq$ 0.3
% \end{itemize}

% Table \ref{tab:parameter11} lists down the variables, their average RMS value (column 12) for template fits, assigned quality flags (column13), and if they are blazhko or not (column 14). We find that in total 30 out of 64 RRLs in the sample have flags A, while 18 have flag B, leaving just 16 with the flag C. 

% Figures \ref{img:temp1to10} \, to \ref{img:tempv71to92}, show the template fits all the RRab and RRc variables and the final template fit for them.

A histogram illustrating the distribution of RMS values for the template fits around the observed light curves is presented in Figure \ref{img:rmsscatter}. The majority of variables exhibit a mean RMS below 0.1. Based on visual inspection of the template fits and insights from the mean RMS histogram, we categorized the light curves into three quality flags:  

\begin{itemize}  
    \item \textbf{A}: Well-defined light curves in all filters, resulting in high-quality template fits (0 $<$ RMS $\leq$ 0.1).  
    \item \textbf{B}: Noticeable scatter in light curves, but clear periodicity and reasonably good template fits (0.1 $<$ RMS $\leq$ 0.2).  
    \item \textbf{C}: Significant scatter in light curves, leading to poor template fits (0.2 $<$ RMS $\leq$ 0.3).  
\end{itemize}  

Table \ref{tab:parameter11} provides an overview of the variables, listing their average RMS values (column 12), assigned quality flags (column 13), and Blazhko classification (column 14). Out of the 64 RR Lyrae stars analyzed, 30 received a quality flag of A, 18 were classified as B, and 16 were assigned a flag C.  

Figures \ref{img:temp1to10} to \ref{img:tempv71to92} display the template fits for all RRab and RRc variables, showcasing their final fitted light curves.

% \section{Results} \label{sec:results}
% \subsection{Period Distribution} \label{sec-perioddist}
% We plotted a histogram in figure \ref{img:perdist} to visualise the period distribution of RRab and RRc variables in the cluster and found their respective means to agree well with the classification of M53 as an Oosterhoff Type II (OoII) cluster. We found mean periods of 0.649 days for the RRab stars (${\rm <}P_{ab}{\rm>}$) in M53, and of 0.3463 days for RRc stars (${\rm<}P_{c}{\rm>}$) in M53. The population of RRc stars in M53 (one of the highest) is at a ratio of 0.547, which again agrees with the value of this parameter in OoII clusters. 

% We can determine the trend in variation of physical parameters in a cluster through the knowledge of period distribution due to the linear dependence of these properties on the periods of the RRLs. Since most properties like metallicity, temperature, and mass vary linearly with the period, period distribution is an indirect source of understanding the distribution of mass, temperature, and metals, in the population of RRLs in a cluster \citep{castellani_rr_2003}.
% \begin{figure}
%     \centering
%     \includegraphics[width=\linewidth]{images/Period Distribution 1.pdf}
%     \caption{Histogram of RRc (cyan) and RRab (red) periods in M53. The mean period for RRab is 0.649 days. This along with the high RRc ratio ($N_c/N_{tot}$) of 0.55 makes M53 a clear Oosterhoff type II cluster. The mean period for RRc is 0.346 days.}
%     \label{img:perdist}
% \end{figure}

\section{Results} \label{sec:results}
\subsection{Period Distribution} \label{sec-perioddist}

\begin{figure}
    \centering
    \includegraphics[width=\linewidth]{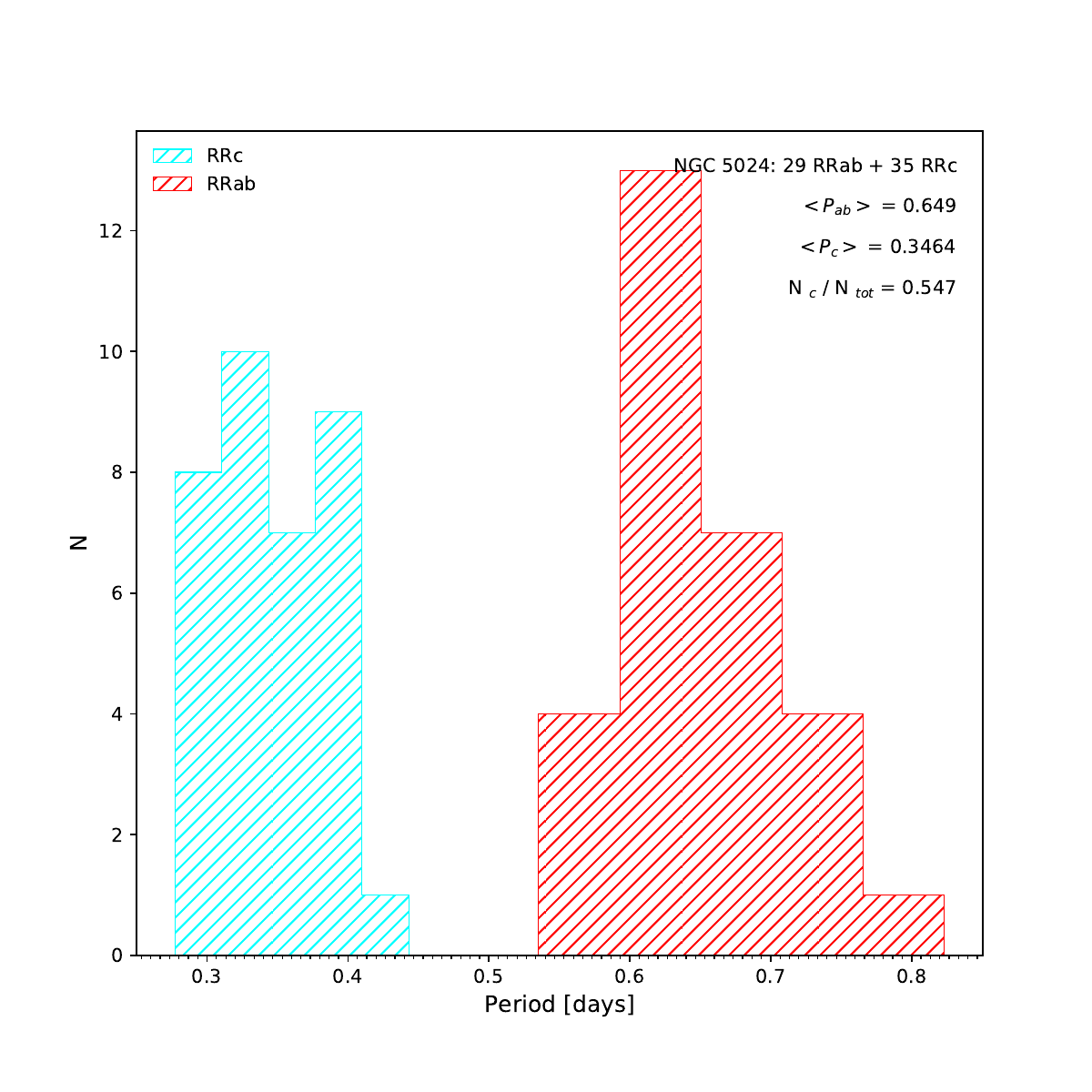}
    % \caption{Histogram of RRc (cyan) and RRab (red) periods in M53. The mean period for RRab is 0.649 days. This, along with the high RRc ratio ($N_c/N_{tot}$) of 0.55, classifies M53 as an Oosterhoff type II cluster. The mean period for RRc is 0.346 days.}
    \caption{Histogram of RRc (cyan) and RRab (red) periods in M53. The mean RRab period is 0.649 days, and the mean RRc period is 0.346 days. Combined with the high RRc fraction ($N_c/N_{\text tot} = 0.55$), this classifies M53 as an Oosterhoff type II cluster.}
    \label{img:perdist}
\end{figure}

To visualize the period distribution of RRab and RRc variables in M53, we plotted a histogram (Figure \ref{img:perdist}). The mean periods of RRab and RRc stars were found to be 0.649 days (${\rm <}P_{ab}{\rm>}$) and 0.3463 days (${\rm <}P_{c}{\rm>}$), respectively, which align well with the classification of M53 as an Oosterhoff Type II (OoII) cluster. Additionally, M53 exhibits a high RRc population ratio of 0.547, further supporting its OoII classification.

Period distribution provides insights into the variation of physical parameters within the cluster, as many stellar properties-such as metallicity, temperature, and mass-exhibit a linear dependence on period. Consequently, analyzing period distribution helps infer the overall distribution of these parameters in the RR Lyrae population of M53 \citep{castellani_rr_2003}.

\subsection{Color-Magnitude Diagram} \label{sec:colormagnitude}

\begin{figure*}
    \centering
    \includegraphics[width=\textwidth]{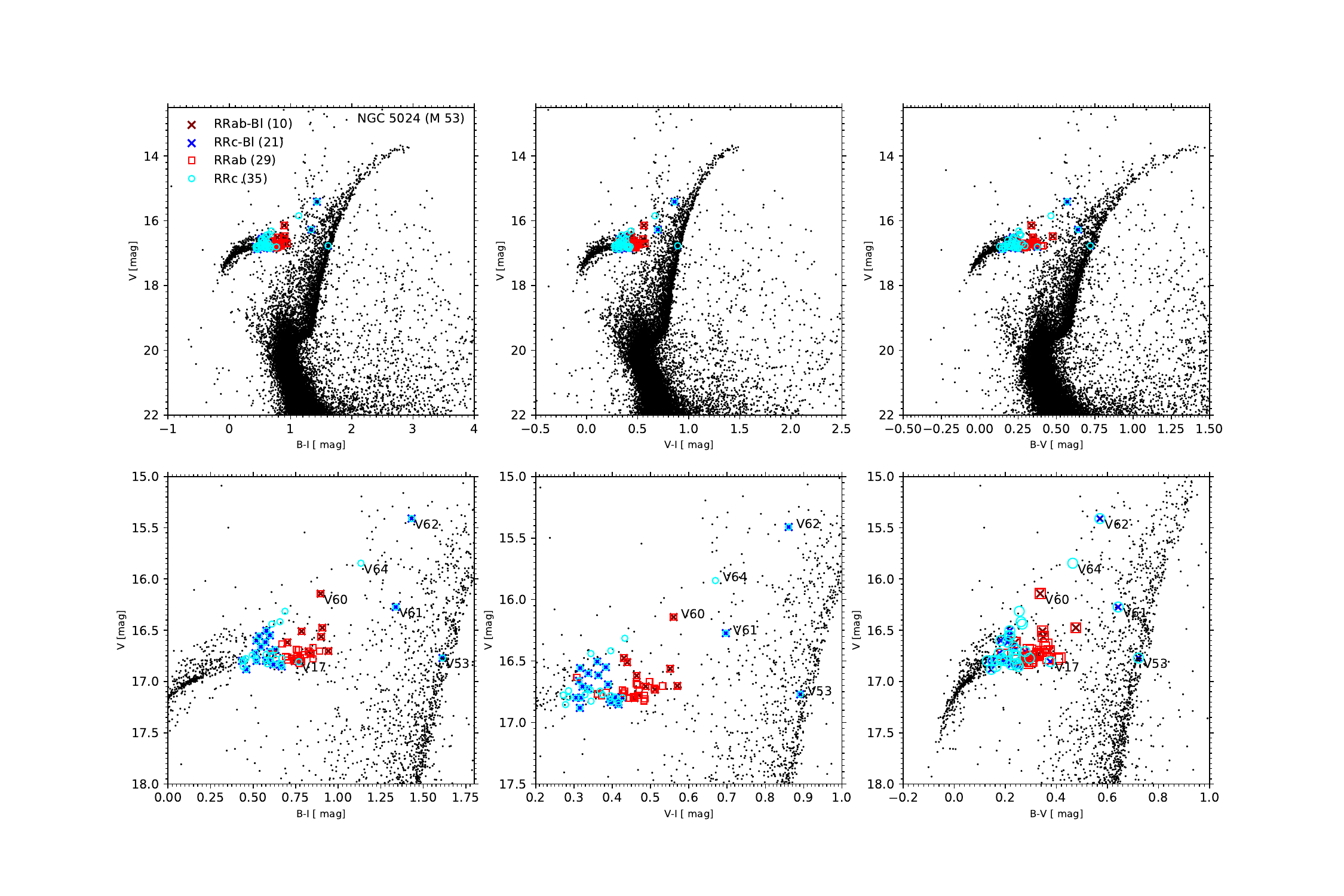} 
    % \caption{Color-Magnitude Diagrams for the cluster M53. The top row presents three plots: B-I vs. V (left), V-I vs. V (middle), and B-V vs. V (right). Cyan circles represent RRc stars, red squares indicate RRab stars, blue crosses mark RRc Blazhko stars, and maroon crosses represent RRab Blazhko stars. The bottom row provides zoomed-in views of the Horizontal Branch (HB) region for each corresponding CMD.}
    \caption{Color–Magnitude Diagrams (CMDs) of the cluster M53. The top row shows B–I vs. V (left), V–I vs. V (middle), and B–V vs. V (right). RRc stars are shown as cyan circles, RRab stars as red squares, RRc Blazhko stars as blue crosses, and RRab Blazhko stars as maroon crosses. The bottom row displays zoomed-in views of the Horizontal Branch (HB) region for the corresponding CMDs.}

    \label{img:colormag}
\end{figure*}

We plotted the V-band magnitudes of RRLs against three colors (B-I, V-I, and B-V) in Figure \ref{img:colormag}. The results indicate that the RRLs lie on the locus of the instability strip in all three diagrams. Such plots, known as Color-Magnitude Diagrams (CMDs), provide insight into the evolutionary state of the stars under investigation. Previous studies, such as \cite{cacciari_multicolor_2005}, have shown that bluer colors are more prone to instability due to shock-induced effects in the interior of RRLs. Consequently, longer-wavelength color indices like V-I and B-I are preferable for deriving accurate stellar parameters.

To correct the magnitude measurements for interstellar reddening due to extinction, the values were adopted from \citep{sarajedini_rr_2006} as provided in Table \ref{tab:reddratio}. The known extinction value in the B-V color, E(B-V), for M53 is 0.02 \citep{safonova_variables_2011, bhardwaj_rr_2021}. The values of the ratio of total to selective absorption, $\left(\frac{A_{\lambda}}{\rm E}\right)$, for each band $\lambda$ were adopted from Table \ref{tab:reddratio} \citep{schlegel_maps_1998, haschke_new_2011}. As the interstellar medium scatters more light at shorter wavelengths, objects appear redder than their intrinsic color. Equation \ref{eq:redd} provides the corrected magnitude $m_r$ from the observed magnitude $m_{uc}$, incorporating E(B-V) and $\frac{A_{\lambda}}{\rm E}$.

\begin{equation}\label{eq:redd}
    m_r = m_{uc} - \left(\frac{A_{\lambda}}{\rm E}\right) \times {\rm E(B-V)}.
\end{equation}

% \begin{table}
%     \centering
%     \caption{The ratios of selective absorption to total absorption of light of different wavelengths due to ISM.}
%     \begin{tabular}{cc}
%         \hline 
%         Filter ($\lambda$) & $\frac{A_{\lambda}}{\rm E}$ \\
%         \hline 
%         U & 5.43 \\
%         B & 4.32 \\
%         V & 3.32 \\
%         R & 2.67 \\
%         I & 1.94 \\
%         \hline
%     \end{tabular}
%     \label{tab:reddratio}
% \end{table}

\begin{table}
    \centering
    \caption{Ratios of total to selective absorption, $\frac{A_{\lambda}}{\rm E}$, for different photometric filters due to interstellar extinction.}
    \begin{tabular}{cc}
        \hline 
        Filter ($\lambda$) & $\frac{A_{\lambda}}{\rm E}$ \\
        \hline 
        U & 5.43 \\
        B & 4.32 \\
        V & 3.32 \\
        R & 2.67 \\
        I & 1.94 \\
        \hline
    \end{tabular}
    \label{tab:reddratio}
\end{table}

Figure \ref{img:colormag} presents various color-magnitude diagrams plotted for the stars of M53. The top panels display the plots for B-I vs. V (left), V-I vs. V (middle), and B-V vs. V (right). The bottom panel shows zoomed-in plots of the HB, where RRLs are clustered. In each of these plots, the long-period RRab stars are seen clustering toward the redder edge of the HB, while the short-period RRc stars are grouped toward the hotter, bluer edge. Some intermixing is also observed in the transition region between RRc and RRab stars, which is more prominent in the V-I color. In contrast, for the B-I color, only one RRc star (V17) appears farther toward the red edge than the other RRc stars. The positions of RRLs in the CMD have been suggested to provide clues about their evolutionary history to some extent \citep{kunder_rr_2013}.

We observe that in each of the CMD plots, certain variables are located far from the expected HB. These variables include V53, V60, V61, V62, and V64, with V64 being the only one that does not exhibit the Blazhko effect. While V60's color falls within the expected RRab region, its V-band magnitude is higher than the bulk, making it appear as an outlier. All the other outlier variables are significantly distant from the HB, with V53 positioned on the Red Giant Branch. We infer that this unexpected behavior is due to blending in the light curves of variables located in dense cluster regions. Another possible reason for these stars to show such behavior could be due to the existence of a binary companion \citep{li_two_2023}. These stars cannot be included in further analysis, as they consistently appear as outliers in all figures, indicating the unreliability of their light curves.

\subsection{Bailey's Diagram}\label{sec:bailey}

\begin{figure}
    \centering
    \includegraphics[width=\linewidth]{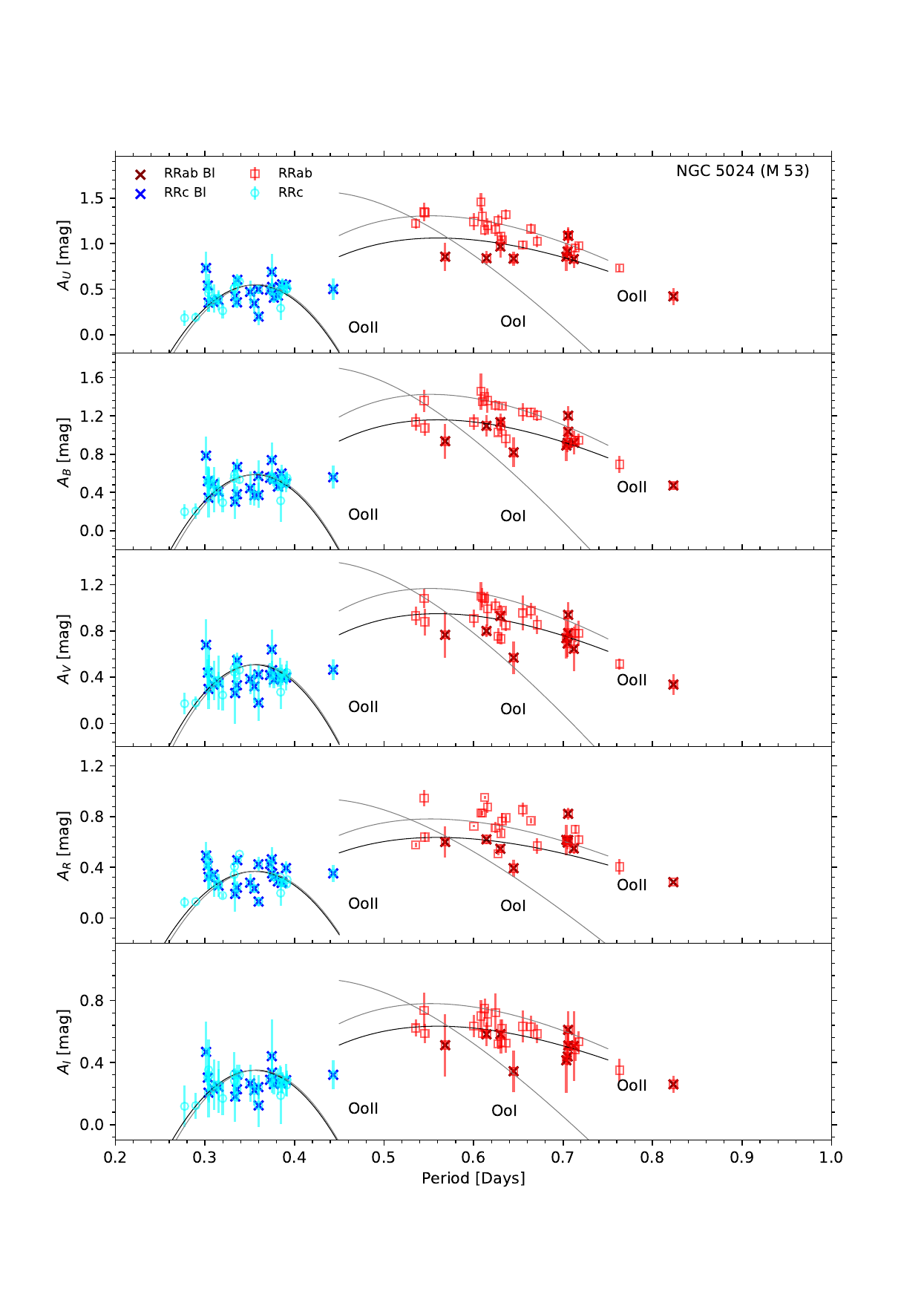}
    % \caption{Bailey diagram for RRLs in M53, where RRc stars are represented by cyan circles and RRab stars by red squares. Blazhko variables among RRc and RRab stars are marked with blue and maroon crosses, respectively. The darker line represents the derived relation for OoII-type clusters for both RRc and RRab. The gray line in the RRc region shows the relation for OoII clusters from \citet{kunder_rr_2013} in the V band, while the OoII relation for RRab (gray) is obtained in the B band from \citet{cacciari_multicolor_2005}. The relation for OoII-type clusters in the I band for RRab (gray) is taken from \citet{arellano_ferro_exploring_2011}. The relations from \citet{cacciari_multicolor_2005} and \citet{arellano_ferro_exploring_2011} were used to fit the dark gray relations to M53. The corresponding relations in other bands were obtained by scaling using the amplitude ratios derived in Section \ref{sec:ampratio}.}
    \caption{Bailey diagram for RRLs in M53. RRc stars are shown as cyan circles and RRab stars as red squares, with Blazhko variables marked by blue (RRc) and maroon (RRab) crosses. The dark line represents the derived OoII-type relation for both RRc and RRab stars. In the RRc region, the gray line corresponds to the OoII relation in the $V$ band from \citet{kunder_rr_2013}. For RRab stars, the gray relations are taken from the $B$ band \citep{cacciari_multicolor_2005} and the $I$ band \citep{arellano_ferro_exploring_2011}. The dark gray fits to M53 were obtained using the \citet{cacciari_multicolor_2005} and \citet{arellano_ferro_exploring_2011} relations, with the corresponding relations in other bands scaled using amplitude ratios derived in Section~\ref{sec:ampratio}.}

    \label{img:bailey}
\end{figure}

We plotted the Bailey diagrams (amplitude vs. period) for the RRLs of M53 in Figure \ref{img:bailey}, which clearly distinguish between RRc and RRab stars, as they occupy different regions of the diagram. By definition, amplitudes are independent of reddening and distance and are also free from zero-point errors \citep{braga_rr_2016}. Stars of the same subtype follow a common period-amplitude relation, which is quadratic in nature, as also noted by \cite{braga_rr_2016}.

We verified the validity of period-amplitude relations from previous studies by scaling them across different bands using amplitude ratios and over-plotting them in gray on the Bailey diagrams for each band. The relation for OoII RRc was obtained from Equation 4 of \cite{kunder_rr_2013} in the V band. The relation for OoII RRab was taken from \cite{arellano_ferro_exploring_2011} in the I band, while the relation for OoI RRab was derived from \cite{cacciari_multicolor_2005} in the B band. These relations were then scaled to other bands using the amplitude ratios obtained in Section \ref{sec:ampratio}. The quadratic relations from \cite{kunder_rr_2013} and \cite{arellano_ferro_exploring_2011} were used as references to fit curves to the actual M53 data, which are shown as darker lines in both the RRc and RRab regions. We observe that in the RRc region, the derived curve aligns very well with the theoretical curve.

We note that both RRc and RRab stars follow their expected amplitude trends, with RRc stars exhibiting a near hairpin variation and RRab stars showing a gradual decrease in amplitude with increasing period. RRab stars are expected to display a gradual amplitude decrease with rising temperature, with a near-flattened peak. This behavior can be explained by the increased efficiency of energy transport in convective regions as stars transition from the hotter to the colder part of the RRab instability strip \citep{bono_pulsation_1994, bono_metal-rich_1997}. RRc stars generally show a nearly constant amplitude with increasing period, although some studies have suggested that RRc stars exhibit a hairpin-like variation \citep{braga_rr_2016}. The curves obtained for RRc (both theoretical and fitted) replicate this hairpin behavior, and the RRc stars of M53 align well with the maxima of the curve. On average, amplitudes decrease as one moves toward longer wavelengths. The period-amplitude relationship for RRab stars shows the least dispersion in the I band, while for RRc stars, it appears relatively consistent across all bands.

The relations obtained for RRab and RRc variables in the I band and V band, respectively for M53, are as follows:

\begin{align}\label{eq:rrabbailey}
    {\rm RRab:} \quad A_I &= -0.221 - 6.786 \log P_{\rm ab} - 13.473 (\log P_{\rm ab})^2,
\end{align}

\begin{align}\label{eq:rrcbailey}
    {\rm RRc:} \quad A_V &= -9.220 + 54.735 P_{\rm c} - 77 P^2_{\rm c}.
\end{align}

We find the standard deviations of the data points around the obtained relations for RRc and RRab to be $0.158$ and $0.079$, respectively. The fact that the error in the coefficient of $P^2$ in Equation \ref{eq:rrcbailey} is of the same magnitude as the coefficient itself highlights the nearly linear distribution of RRc stars in the Bailey diagram, leaving the hairpin distribution still subject to contradictions.

\subsection{Period-Luminosity Relation} \label{sec:periodluminosity}

\begin{figure}
    \centering
    \includegraphics[width=\linewidth]{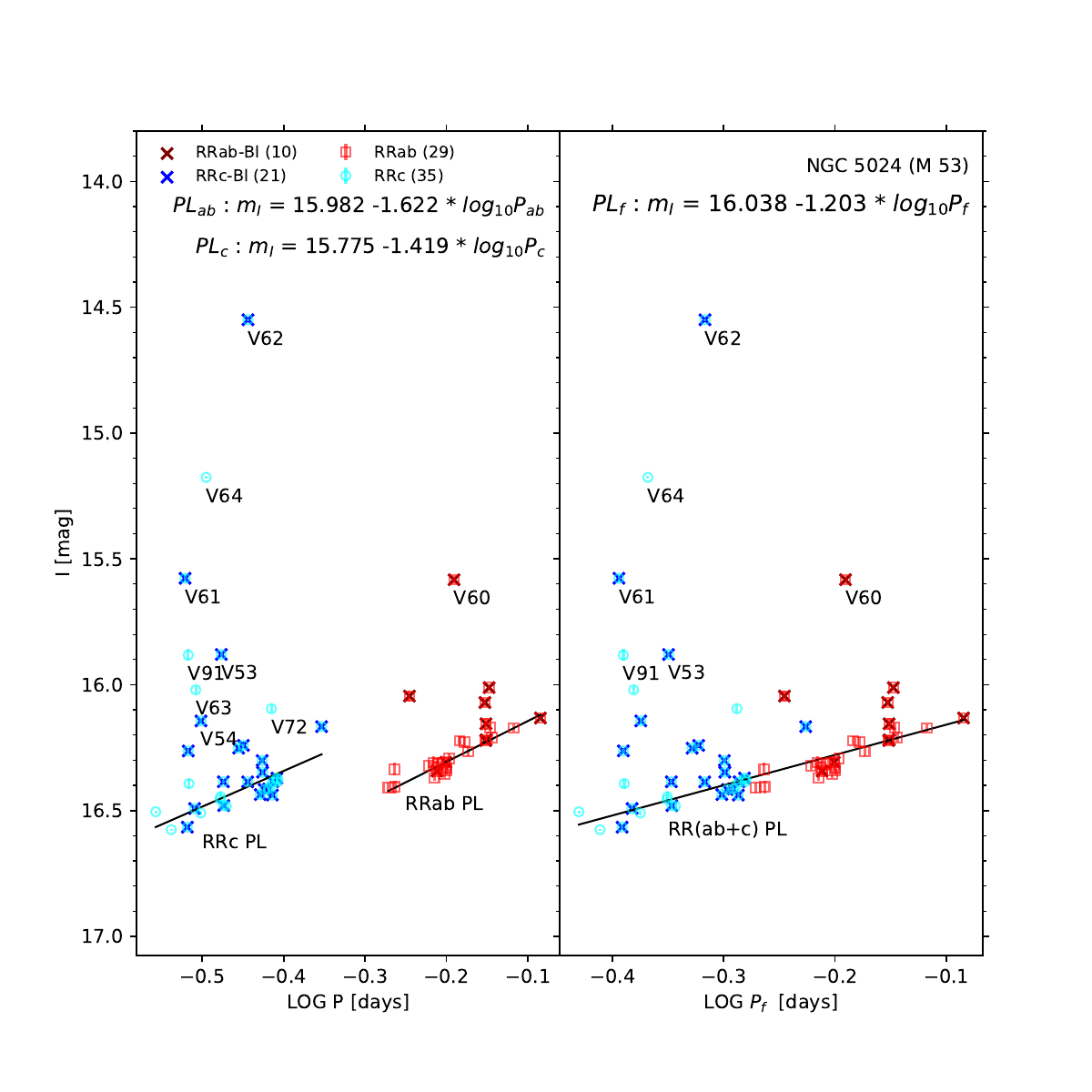}
    % \caption{I-band luminosity vs. logarithm of period (Period-Luminosity) plot for the RRLs in M53. The colored symbols follow the same scheme as in Figure \ref{img:bailey}. The left panel presents the PL relation for RRc and RRab stars separately, while the right panel shows the global PL relation after converting RRc periods to fundamental ones. The relationships are displayed at the top of each panel.}
    \caption{I-band Period–Luminosity (PL) relation for the RRLs in M53. The symbols follow the same color scheme as in Figure~\ref{img:bailey}. The left panel shows the PL relations for RRc and RRab stars separately, while the right panel displays the global PL relation after converting RRc periods to their fundamental-mode equivalents. The best-fit relations are indicated at the top of each panel.}
    \label{img:periodluminosity}
\end{figure}

We plot the period-luminosity (PL) relations for the RRLs in M53 using reddening-corrected magnitudes in the I band, as shown in Figure \ref{img:periodluminosity}. The left panel of the figure presents the PL relations for RRc and RRab stars separately. As expected, RRc stars appear dimmer (but hotter) compared to RRab stars. In addition to the outliers discussed in Section \ref{sec:colormagnitude}, stars such as V91 and V54 exhibit dispersion in their parameters, which may result from errors in the template fitting process or period determination. The right panel of the same figure shows the I-band \textit{global} PL relation for the entire population. To obtain the global relation, the periods of RRc stars, which pulsate in the first overtone (FO) mode ($P_c$), were fundamentalized to $P_f$ using Equation \ref{eq:fundamentalised} so that they could be analyzed together with RRab stars, which pulsate in the fundamental mode \citep{braga_distance_2015}.  

\begin{equation} \label{eq:fundamentalised}
    \log P_f = \log P_c + 0.127.
\end{equation}
The Table \ref{tab:periodluminosity} presents the PL relations for RRab, RRc and the global mode for the I band magnitude. $\sigma$ represents the total scatter in the PL relation.  

Luminosities are known to depend on metallicities in addition to the period. In fact, RRLs follow a more stringent PLZ (period-luminosity-metallicity) relation, which arises due to the near-linear dependence of absolute luminosity on metallicity, expressed as $M_V = a + b[Fe/H]$.  

A PL relation at longer wavelengths, whereas in the U, B, V, and R bands, the dispersion around the obtained PL relation is significantly larger. This dispersion decreases with increasing wavelength \citep{bhardwaj_optical_2021, braga_distance_2015} because, at longer wavelengths, the sensitivity to temperature variations in the instability strip diminishes. Additionally, we note that moving to longer wavelengths results in a less steep PL relation slope. Due to this PL relation, RRLs within a cluster serve as standard candles for determining distances to their respective clusters and nearby galaxies \citep{garofalo_new_2022}. In Section \ref{sec:distdetermine}, we use this property to robustly determine the distance to M53.  

\begin{table}
    \centering
    % \caption{Coefficients, their associated errors, and the standard deviation obtained for the I-band PL relations for RRab, RRc, and the global RRL population of M53. For the global PL relation, the periods of RRc stars were fundamentalized using Equation \ref{eq:fundamentalised}.}
    \caption{This table presents the coefficients with their uncertainties and the standard deviation of the I-band PL relations for RRab, RRc, and the global RRL population in M53. For the global PL relation, the RRc periods were fundamentalized using Equation \ref{eq:fundamentalised}.}
    \begin{tabular}{cccccc}
        \hline 
        \multicolumn{6}{l}{Form of PL equation: $m_I = a + b \log P$}  \\
        \hline 
        Type & $a$ & $\sigma_a$ & $b$ & $\sigma_b$ & $\sigma$ \\
        \hline 
        RRab & 15.982 & 0.022 & -1.622 & 0.121 & 0.157 \\
        RRc  & 15.775 & 0.090 & -1.419 & 0.184 & 0.458 \\
        RRab + RRc & 16.039 & 0.090 & -1.203 & 0.070 & 0.417 \\
        \hline
    \end{tabular}
    \label{tab:periodluminosity}
\end{table}

\subsection{Period-Wesenheit Relation} \label{sec:periodwesenheit}

The Wesenheit magnitudes, denoted as $W(X, Y-Z)$, are constructed using magnitudes in different filters, where $X$ is the primary band, and $Y$ and $Z$ define the color index. These are computed using the following formula \citep{braga_rr_2016}:

\begin{equation} \label{eq:wesenheit}
    W(X,Y-Z) = X + \frac{A_X}{A_Y-A_Z} \cdot (Y-Z).
\end{equation}

Here, $A_X$, $A_Y$, and $A_Z$ are the respective selective absorption coefficients derived from the reddening law (see Table \ref{tab:reddratio}). Wesenheit indices can be either dual-band or triple-band, depending on whether the same or different filters are used for $X$ and $Z$. The Wesenheit index serves as an intrinsic magnitude that minimizes the scatter in period-luminosity relations, making it a robust tool for standard candle applications \citep{braga_rr_2016}.

We plotted the Period-Wesenheit (PW) relations for RRLs in M53 by calculating Wesenheit magnitudes in three dual-band and one triple-band combination. Unlike PL relations, PW relations have the advantage of being reddening independent by construction, as they incorporate color terms that effectively cancel out extinction effects. This makes them particularly useful for precise distance determinations and comparative studies across different environments.

\begin{figure}
    \centering
    \includegraphics[width=\linewidth]{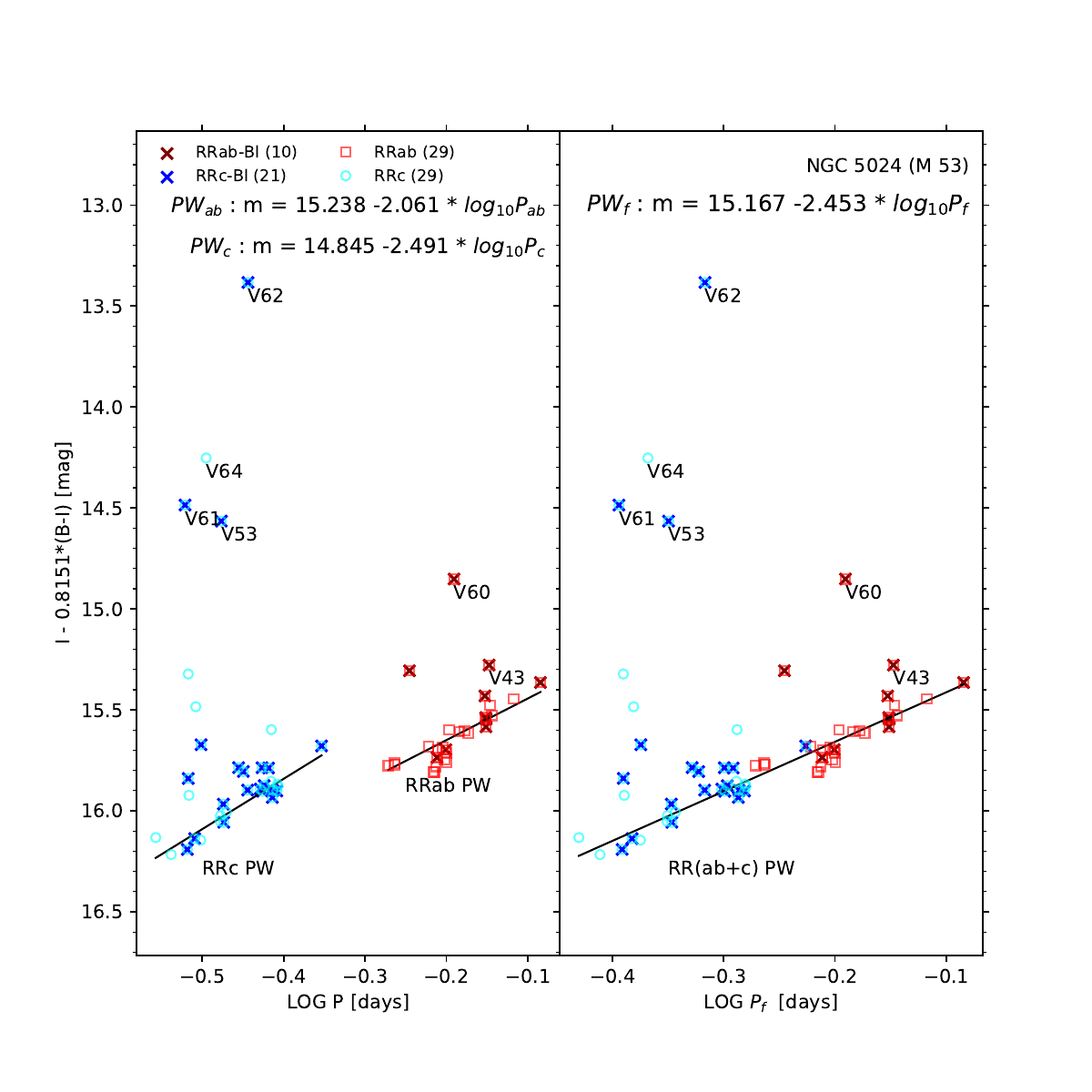}
    \caption{Wesenheit magnitude $W(I,B-I)$ vs. $\log P$ (Period-Wesenheit) plot of the RRLs in M53. The colored symbols follow the same scheme as Figure \ref{img:bailey}. The left panel includes PW relationships for RRc and RRab separately, while the right panel shows the global PW relationship after fundamentalizing the RRc periods. The relationships are displayed at the top of each panel.}
    \label{img:wesenheitIBI}
\end{figure}

\begin{figure}
    \centering
    \includegraphics[width=\linewidth]{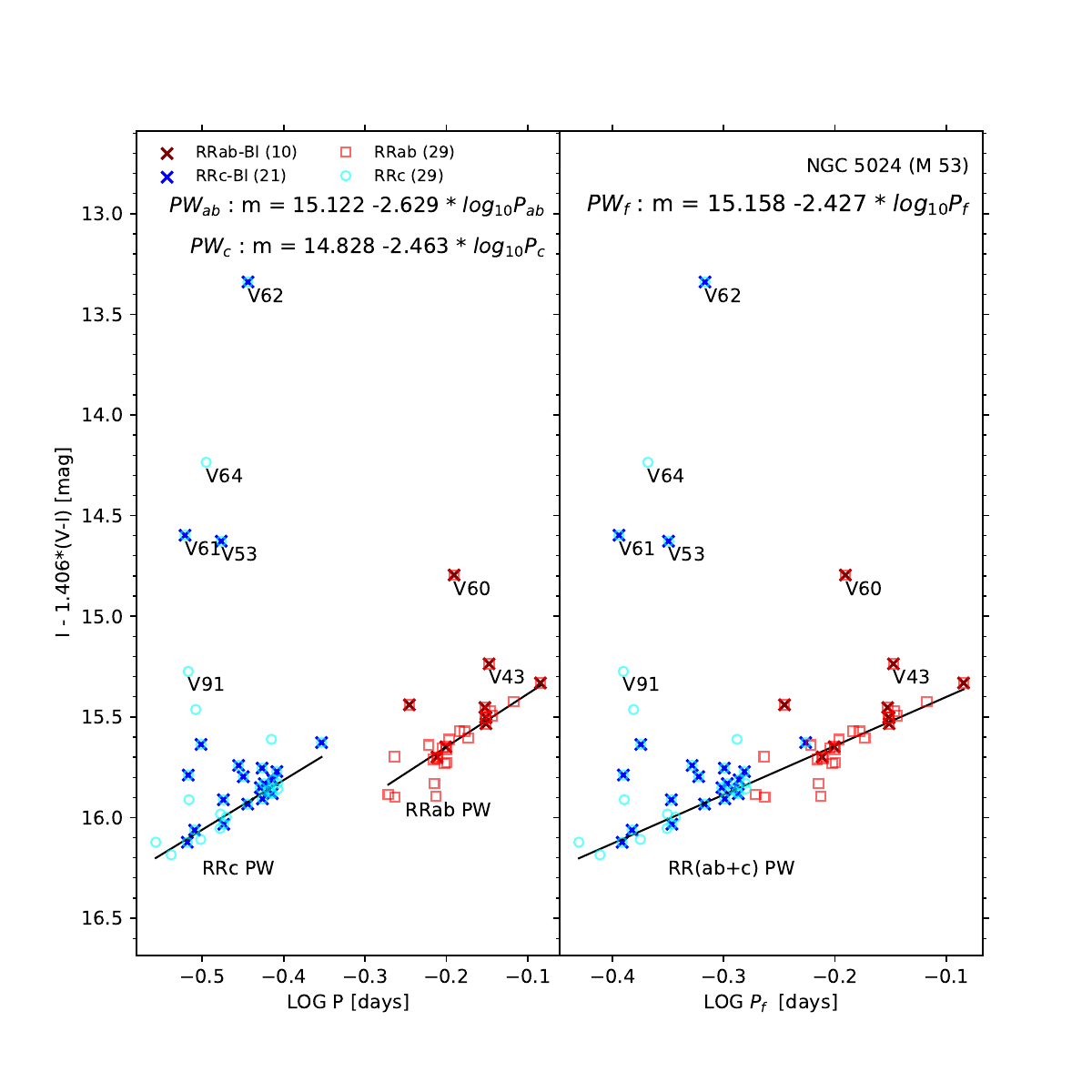}
    \caption{Wesenheit magnitude $W(I,V-I)$ vs. $\log P$ (Period-Wesenheit) plot of the RRLs in M53. The colored symbols follow the same scheme as Figure \ref{img:bailey}. The left panel includes PW relationships for RRc and RRab separately, while the right panel shows the global PW relationship after fundamentalizing the RRc periods. The relationships are displayed at the top of each panel.}
    \label{img:wesenheitIVI}
\end{figure}

\begin{figure}
    \centering
    \includegraphics[width=\linewidth]{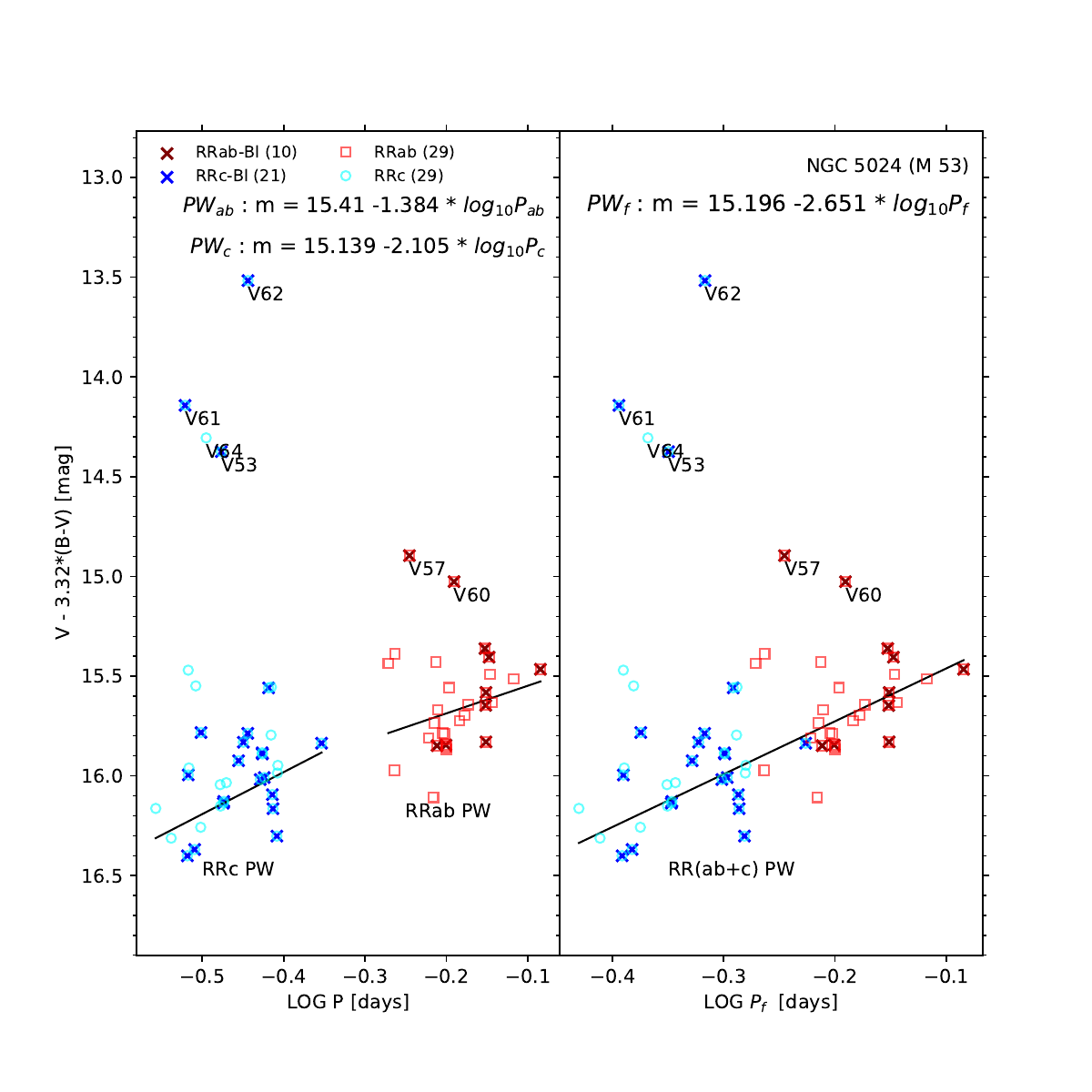}
    \caption{Wesenheit magnitude $W(V,B-V)$ vs. $\log P$ (Period-Wesenheit) plot of the RRLs in M53. The colored symbols follow the same scheme as Figure \ref{img:bailey}. The left panel includes PW relationships for RRc and RRab separately, while the right panel shows the global PW relationship after fundamentalizing the RRc periods. The relationships are displayed at the top of each panel.}
    \label{img:wesenheitVBV}
\end{figure}

\begin{figure}
    \centering
    \includegraphics[width=\linewidth]{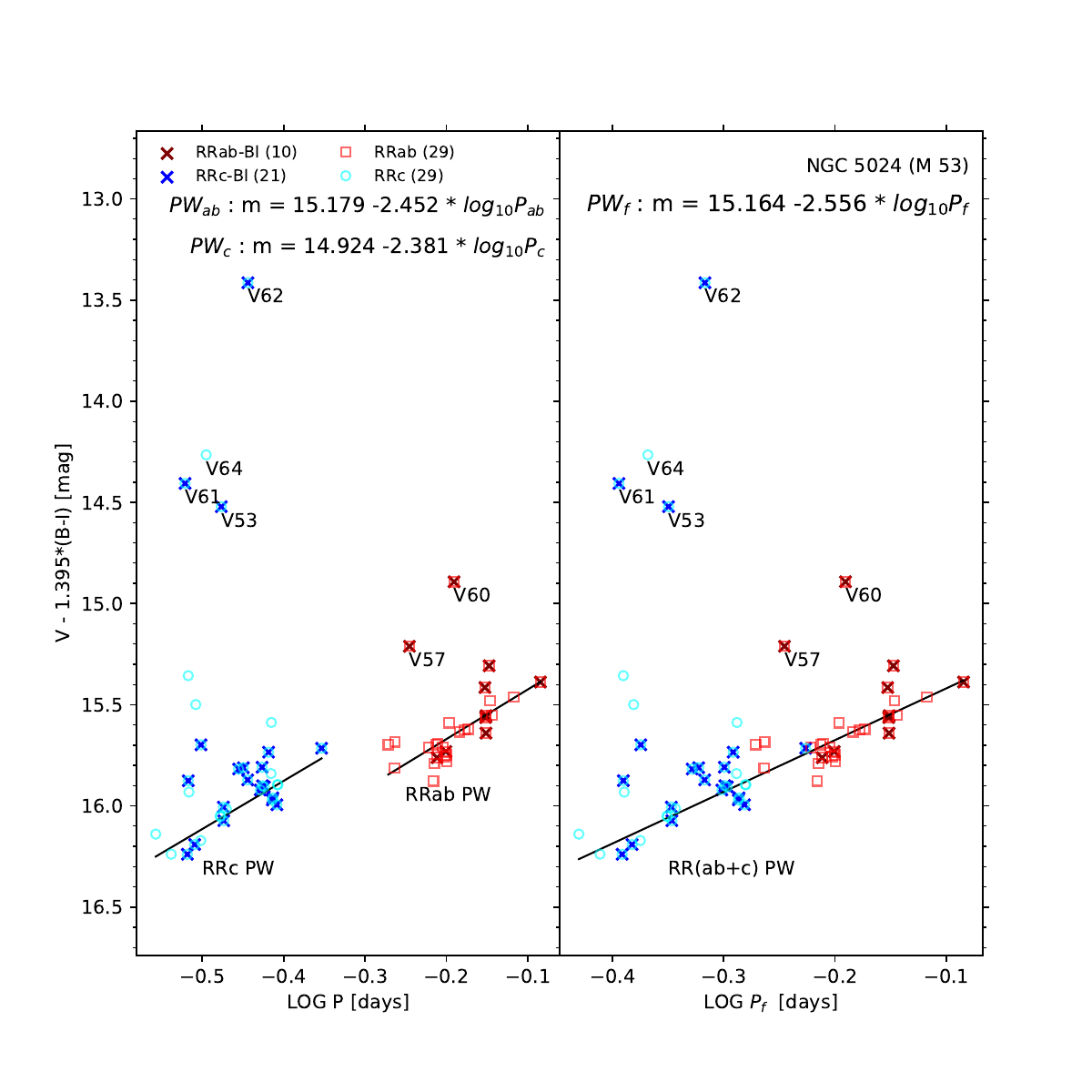}
    \caption{Wesenheit magnitude $W(V,B-I)$ vs. $\log P$ (Period-Wesenheit) plot of the RRLs in M53. The colored symbols follow the same scheme as Figure \ref{img:bailey}. The left panel includes PW relationships for RRc and RRab separately, while the right panel shows the global PW relationship after fundamentalizing the RRc periods. The relationships are displayed at the top of each panel.}
    \label{img:wesenheitVBI}
\end{figure}

Figures \ref{img:wesenheitIBI} to \ref{img:wesenheitVBI} show the PW relations in various band combinations. The left panel in each figure represents the PW relationship for RRc and RRab stars separately, while the right panel shows the global PW relationship after fundamentalizing the RRc periods. The fundamentalization of RRc periods allows both types of RRLs to be analyzed together in a single relation, improving the robustness of distance estimates.

The plots for the four Period-Wesenheit (PW) relations-three dual-band and one triple-band-are shown in Figures \ref{img:wesenheitIBI}, \ref{img:wesenheitIVI}, \ref{img:wesenheitVBV}, and \ref{img:wesenheitVBI}, respectively. The corresponding values of $\frac{A_X}{A_Y-A_Z}$ for each case are 0.8151, 1.406, 3.32, and 1.395 which were adopted from \cite{marconi_new_2015}. 

As seen in the PW plots, variables affected by blending lie far from the main trend. However, ignoring these outliers, the other variables exhibit better-constrained relations, particularly for $W(I,B-I)$ and $W(I,V-I)$. In contrast, $W(V,B-V)$ shows greater dispersion, likely due to the inherent inadequacy of PL relations in the $V$ band, as discussed in Section \ref{sec:periodluminosity}. 

In all cases, RRab stars appear less scattered compared to RRc stars, which may be attributed to the prevalence of Blazhko modulation among RRc variables. The left panel of each PW figure presents the PW relation for RRab and RRc separately after removing outliers. The right panel shows the global PW relation, obtained by fundamentalizing RRc periods using Equation \ref{eq:fundamentalised} and treating them alongside RRab stars.

\begin{table*}
    \centering
    \caption{This table reports the coefficients, their associated errors, and the standard deviation for the PW relations of RRab, RRc, and the combined RRL population in M53. The values of $\sigma$ are calculated after removing distant outliers, which were excluded from the analysis.}
    \begin{tabular}{ccccccc}
        \hline \multicolumn{7}{c}{Form of PW equation: $m = W(X,Y-Z) = a + b\log_{10}P$}  \\
        \hline Type & $\frac{A_X}{A_Y-A_Z}$ & $a$ & $a_{\rm err}$ & $b$ & $b_{\rm err}$ & $\sigma$ \\
        \hline \multicolumn{7}{c}{$W(I,B-I)$}\\
        \hline RRab & 0.8151 & 15.238 & 0.094 & -2.061 & 0.476 & 0.104 \\
        RRc & 0.8151 & 14.845 & 0.085 & -2.491 & 0.186 & 0.041 \\
        RRab + RRc & 0.8151 & 15.167 & 0.085 & -2.453 & 0.151 & 0.084 \\
        \hline \multicolumn{7}{c}{$W(I,V-I)$}\\
        \hline RRab & 1.406 & 15.122 & 0.082 & -2.630 & 0.419 & 0.092 \\
        RRc & 1.406 & 14.828 & 0.094 & -2.463 & 0.203 & 0.041 \\
        RRab + RRc & 1.406 & 15.158 & 0.094 & -2.427 & 0.138 & 0.076 \\
        \hline \multicolumn{7}{c}{$W(V,B-V)$}\\
        \hline RRab & 3.32 & 15.410 & 0.161 & -1.384 & 0.837 & 0.181 \\
        RRc & 3.32 & 15.139 & 0.254 & -2.105 & 0.554 & 0.120 \\
        RRab + RRc & 3.32 & 15.197 & 0.254 & -2.651 & 0.292 & 0.166 \\
        \hline \multicolumn{7}{c}{$W(V,B-I)$}\\
        \hline RRab & 1.395 & 15.180 & 0.077 & -2.452 & 0.402 & 0.087 \\
        RRc & 1.395 & 14.924 & 0.120 & -2.381 & 0.261 & 0.056 \\
        RRab + RRc & 1.395 & 15.164 & 0.120 & -2.556 & 0.134 & 0.076 \\
        \hline
    \end{tabular}
    \label{tab:periodwesenheit}
\end{table*}

\subsection{Determination of Distance}\label{sec:distdetermine}

We calculated distances to the RRLs of M53 using the method of \cite{braga_rr_2016}, which utilizes PW and PL relations across various bands to derive a robust weighted mean distance to the cluster. RRab and RRc stars are known to follow PL relations, where their brightness varies linearly with period. Since all RRLs adhere to similar relationships, their distances can be estimated by measuring their brightness and applying the corresponding PL/PW relations.

The PL and PW relations obtained in this study relate the apparent magnitude in a given band ($m_{\lambda}$) to the period. To compute distances, one also requires the absolute magnitudes in the respective bands ($M_{\lambda}$). The distance modulus $\mu$ is then given by:

\begin{equation} \label{eq:distance}
    \mu = m_{\lambda} - M_{\lambda},
\end{equation}

which leads to the distance calculation:

\begin{equation}
    d = 10^{\left(\frac{\mu+5}{5}\right)} \, {\rm pc}.
\end{equation}

To determine $M_{\lambda}$, we adopt the theoretical PLZ and PWZ relations from \cite{marconi_new_2015}, which introduce a linear dependence on metallicity ([Fe/H]). This dependence leads to increased dispersion in distances for bands with stronger metallicity sensitivity, as the empirical PL and PW relations lack this correction. We use Table 6 of \cite{marconi_new_2015} for the R- and I-band PLZ relation zero points and coefficients, which provide relations for RRab, RRc, and the global RRL population. Distance estimates improve at longer wavelengths, particularly in the NIR, due to reduced sensitivity to interstellar reddening and metallicity effects.

The PWZ relations incorporate Wesenheit magnitudes, which, being extinction-free by construction, yield more accurate distance estimates. We used Tables 7 and 8 from \cite{marconi_new_2015} to obtain the zero points and coefficients for the theoretical PWZ relations. Among these, the (V, B-I) PWZ relation exhibits the weakest dependence on metallicity and thus provides the most reliable distance estimates. The general form of the PLZ and PWZ relations is:

\begin{equation}\label{eq:plpwform}
\text{magnitude} = a + b \log P + c [\text{Fe/H}].
\end{equation}

For M53, we adopted [Fe/H] = -2.06 dex from \cite{bhardwaj_rr_2021}. The PWZ and PLZ relation plots are shown in Figures \ref{img:muvbi}, \ref{img:muvbv}, \ref{img:muibi}, \ref{img:muivi}, \ref{img:mui}, and \ref{img:mur}. The left panels of these figures display the calculated $\mu$ values for RRab and RRc stars separately (excluding the global sample for clarity), while the right panels show histograms for all three categories.

Notably, several outliers appear in the left panels. All CMD outliers also deviate in these plots, along with additional outliers-typically V63, V91, V57, and V54. The CMD outliers are excluded from the histograms and weighted mean calculations (shown at the bottom of each left panel) due to their confirmed contamination (see Section \ref{sec:colormagnitude}). However, the additional outliers are retained in the analysis.

Examining the histograms, the I-band PLZ relation produces more consistent distances compared to the R-band PLZ. Among PWZ relations, (V, B-V) exhibits the highest dispersion, whereas (V, B-I) and (I, B-I) show the least scatter, making them the most reliable. Given the reduced metallicity dependence of (V, B-I), we adopt the distance modulus $\mu$ derived from this PWZ relation. The weighted mean distance moduli for each case (RRc, RRab, Global) are listed in Table \ref{tab:wmeanmu}.

\begin{table}
    \centering
    \caption{Weighted mean distance moduli ($\langle\mu\rangle$) calculated using various PWZ and PLZ relations for RRc, RRab, and the global RRL population in M53.}    
    \begin{tabular}{cccc}
         \hline Band & RRc & RRab & Global  \\
         \hline &\multicolumn{3}{c}{$\langle\mu\rangle$ [mag]}\\
         \hline V,B-I & 16.246 & 16.273 & 16.256 \\ 
         V,B-V & 16.231 & 16.135 & 16.188 \\
         I,B-I & 16.261 & 16.324 & 16.274 \\
         I,V-I & 16.272 & 16.372 & 16.310 \\
         I & 16.237 & 16.340 & 16.285 \\
         R & 16.218 & 16.301 & 16.267 \\
         \hline
    \end{tabular}
    \label{tab:wmeanmu}
\end{table}

\section{Discussion} \label{sec:discussion}

\subsection{Oosterhoff Classification}
The Oosterhoff classification, introduced by \cite{oosterhoff_remarks_1939}, distinguishes globular clusters (GCs) based on the properties of their RR Lyrae stars. Two primary groups, Oosterhoff I (OoI) and Oosterhoff II (OoII), have been identified, with a third, more metal-rich subclass suggested by \cite{braga_rr_2016}.

OoI clusters exhibit higher metallicity, an average RRab period of 0.55 days, and a fundamental mode (FO) to total RRL ratio of approximately 0.17. Notable examples include \(\omega\) Centauri \citep{braga_rr_2016} and M3 \citep{kumar_multiwavelength_2024}. In contrast, OoII clusters, which include M53, are more metal-poor, with longer mean RRab periods (\(\sim\)0.65 days) and a higher FO-to-total ratio of 0.47 \citep{castellani_rr_2003}. The proposed OoIII subclass features even longer RRab periods ($\sim 0.74$ days) and includes NGC 6388 and NGC 6441.

The origin of the Oosterhoff effect remains debated. A commonly proposed explanation attributes it to metallicity variations \citep{molnar_first_2021}, where decreasing metallicity results in longer pulsation periods. Additionally, horizontal branch (HB) morphology plays a significant role. Interestingly, OoII clusters in the Milky Way appear to be spatially correlated \citep{harris_spatial_1976}, suggesting a common extragalactic origin. This challenges the earlier hypothesis that OoII clusters are the oldest in the Milky Way.

Unlike in globular clusters, the Oosterhoff dichotomy is absent among field RRLs, which exhibit a continuous metallicity distribution \citep{fabrizio_use_2021}. Furthermore, studies have found no clear Oosterhoff classification in dwarf galaxies of the Local Group \citep{sesar_light_2009}.

M53 firmly classifies as an OoII cluster, as evidenced by its period distribution (Figure \ref{img:perdist}), with an average RRab period of 0.649 days and an RRc-to-total RRL ratio of 0.547. The Bailey diagram (Figure \ref{img:bailey}) further supports this classification, showing that RRab stars align with OoII loci, and RRc stars follow the characteristic OoII ``hairpin" distribution.

\subsection{Outlier Variables}
Several variables, namely V53, V60, V61, V62, V64, and possibly V17, exhibit anomalous behavior in multiple diagnostic plots. V17 appears significantly displaced in the B-I vs. V color-magnitude diagram (CMD; Figure \ref{img:colormag}), suggesting contamination from a nearby star in the dense cluster core. The remaining stars display outlier behavior across period-luminosity (PL) and period-Wesenheit (PW) relations, distance moduli, metallicity estimates, and absolute magnitude calculations. Due to their systematic deviations, these variables were excluded from weighted mean calculations.

These variables also presented challenges in period determination, exhibiting irregular light curves. During period fragmentation (Section \ref{sec:period_problematic}), significant portions of their data were ignored. Consequently, in analyses that rely strictly on period and amplitude, such as luminosity and effective temperature plots, these variables do not deviate significantly.

Additionally, V54, V91, and V57 appear as occasional outliers in distance calculations (Section \ref{sec:distdetermine}). Unlike the previously mentioned outliers, these variables were retained in weighted mean calculations, as their light curve irregularities did not systematically skew pulsation parameter determination.

\subsection{The Blazhko Effect in M53} \label{sec-blazhko}
The Blazhko effect, first identified by \cite{blazhko_mitteilung_1907}, describes long-period modulations in RR Lyrae pulsation properties, typically occurring over 10 to 100 days. While primarily observed in fundamental-mode (FU) RR Lyrae, it also affects first-overtone (FO) stars.

Despite various proposed explanations, no single theory has achieved consensus. One prevalent hypothesis attributes the Blazhko effect to nonlinear mode resonances \citep{molnar_first_2021, jameson_rr_1986, hoffman_periods_2021}, particularly involving the sixth or ninth harmonic of the fundamental pulsation frequency \citep{plachy_rr_2021}.

Blazhko modulation complicates the determination of fundamental stellar parameters, leading to uncertainties in period determination, distance estimates, metallicity calculations, luminosities, masses, and effective temperatures. M53 is notable for its high incidence of Blazhko RRc stars, with 21 of 35 RRc variables exhibiting modulation. Among RRab stars, 10 out of 29 show Blazhko-like behavior. Furthermore, 11 of the 19 variables requiring segmented period determination were identified as Blazhko stars. Blazhko variables frequently appear as outliers in diagnostic plots and show higher photometric noise, complicating template fitting procedures.

\subsection{Distances}
% Recent distance modulus estimates for M53, compiled in Table 3 of \cite{bhardwaj_rr_2021}, range from 16.3 to 16.5 mag, corresponding to a distance of .... Kpc. The same study adopts a distance modulus of 16.403 mag as the reference value.

In this study, distance moduli were computed using PLZ and PWZ relations alongside Fourier parameters. The resulting weighted mean values span from 16.18 mag (V, B-V: PWZ Global) to 16.37 mag (I, V-I: PWZ RRc), aligning well with literature estimates. 
% Figure \ref{img:mucompare} illustrates the deviation of these values from the reference modulus of 16.403 mag. Error bars exclude CMD outliers (Section \ref{sec:colormagnitude}), with uncertainties reflecting the difference between our calculations and the adopted standard value.

The results demonstrate that PLZ/PWZ relations provide robust distance estimates, with PWZ relations generally yielding lower dispersion due to their reduced sensitivity to interstellar extinction. Among these, the (V, B-I) PWZ relation is the most reliable, as discussed in Section \ref{sec:distdetermine}.

The weighted mean distance modulus for NGC 5024 was found to be $16.242 \pm 0.05$ mag, corresponding to a distance of $17.72 \pm 0.41$ Kpc. This result is in excellent agreement with the findings of \cite{muraveva_metallicity_2024}, who determined a distance modulus of $16.27 \pm 0.06$ mag ($17.95 \pm 0.50$ Kpc) for the same cluster using Gaia DR3 photometric data.

However, our result is slightly lower than the value of $16.403 \pm 0.051$ mag ($19.08 \pm 0.45$ Kpc) adopted by \cite{bhardwaj_rr_2021}. The parallax-based distance from \cite{hunt_improving_2024} is significantly higher, at approximately $25.02$ Kpc. Due to the very large uncertainty in their parallax measurement ($0.03997370 \pm 0.06506437$ mas; from Gaia DR3 parallax data), this value is considered less reliable than the distance modulus estimates. Our findings, therefore, contribute to a more accurate and consistent distance scale for globular clusters.

\section{Conclusions} \label{sec:conclusion}
In this study, we conducted a comprehensive photometric analysis of 64 RR Lyrae stars in the globular cluster NGC 5024 (M53) using UBVRI multiband observations. Our methodological approach, which included a strict data filtering criterion and a segmented Lomb-Scargle period search, allowed for a robust determination of pulsation periods, particularly for RRc stars with complex light curves. By employing template light curve fitting based on mean amplitude ratios, we were able to derive mean magnitudes and amplitudes, even in the presence of photometric scatter and phase gaps. The high quality of our light curve fits is reflected in the quality flags assigned to the variables, with 30 stars receiving the highest ``A" classification.

Our analysis of the pulsation properties of the RR Lyrae population firmly establishes NGC 5024's classification as an Oosterhoff II (OoII) cluster, consistent with its metal-poor nature. We found the mean periods of RRab and RRc stars to be 0.649 and 0.346 days, respectively, and an RRc-to-total ratio of 0.547, which align well with the expected properties of OoII clusters. The analysis of the Bailey diagram further corroborated this classification, with the M53 RR Lyrae stars occupying the expected loci for OoII clusters and yielding well-constrained period-amplitude relations. Specifically, the I-band amplitude-period relation for RRab stars is given by:

\begin{equation}
A_I = -0.221 - 6.786\log_{10}P - 13.473(\log_{10}P)^2
\end{equation}
and the V-band relation for RRc stars is:
\begin{equation}
A_V = -9.220 + 54.735P - 77P^2
\end{equation}

Our investigation of the color-magnitude diagrams revealed that several variables (V53, V60, V61, V62, V64, and possibly V17) lie significantly outside the instability strip. These stars were consistently identified as outliers in our analyses, suggesting their photometry is likely contaminated by nearby stellar sources. We derived highly precise period-luminosity (PL) and period-Wesenheit (PW) relations, utilizing reddening-free Wesenheit magnitudes, which reduced the scatter in our distance estimates. The weighted mean distance modulus for NGC 5024 was found to be $16.242 \pm 0.05$ mag, placing the cluster at a distance of $17.72 \pm 0.41$ Kpc. This result is in excellent agreement with the findings of \cite{muraveva_metallicity_2024}, who determined a distance modulus of $16.27 \pm 0.06$ mag ($17.95 \pm 0.50$ Kpc) using Gaia DR3 photometric data. Our result is slightly lower than the value of $16.403 \pm 0.051$ mag ($19.08 \pm 0.45$ Kpc) adopted by \cite{bhardwaj_rr_2021}, but the differences are small and all are highly consistent with each other. Overall, this study provides a comprehensive characterization of RR Lyrae stars in NGC 5024, reinforcing its classification as an OoII cluster. The refined pulsation parameters, period-luminosity and period-Wesenheit relations, and improved distance estimates contribute valuable insights into the properties of this metal-poor globular cluster.

\section*{Acknowledgements}
We are grateful to P. B. Stetson for generously sharing the optical light curves of RR Lyrae stars in M53 used in this study. N.K. acknowledges the use of the High Performance Cluster facility, Pegasus of IUCAA, Pune, for providing the computational~resources.

% This research has made use of the \textsc{Astropy} package \citep{astropy_2013, astropy_2018}, along with other scientific Python libraries including \textsc{NumPy} and \textsc{Matplotlib}.  

% N.K. acknowledges financial support from [Funding Agency, Grant Number (if applicable)] and is grateful to [Institution/Department] for providing computational resources and research facilities. We also extend our gratitude to our colleagues at [Your Institution] for their helpful discussions and insights.  

% Finally, we appreciate the anonymous referee for their constructive comments, which helped improve the quality of this manuscript.  

%%%%%%%%%%%%%%%%%%%%%%%%%%%%%%%%%%%%%%%%%%%%%%%%%%
\section*{Data Availability}
The data underlying this article, including the full electronic tables, will be available in the article’s online supplementary material.

%%%%%%%%%%%%%%%%%%%%%%%%%%%%%%%%%%%%%%%%%%%%%%%%%%

\appendix
\section{Additional Figures}
\begin{figure*}
    \centering
    \includegraphics[width=0.85\linewidth]{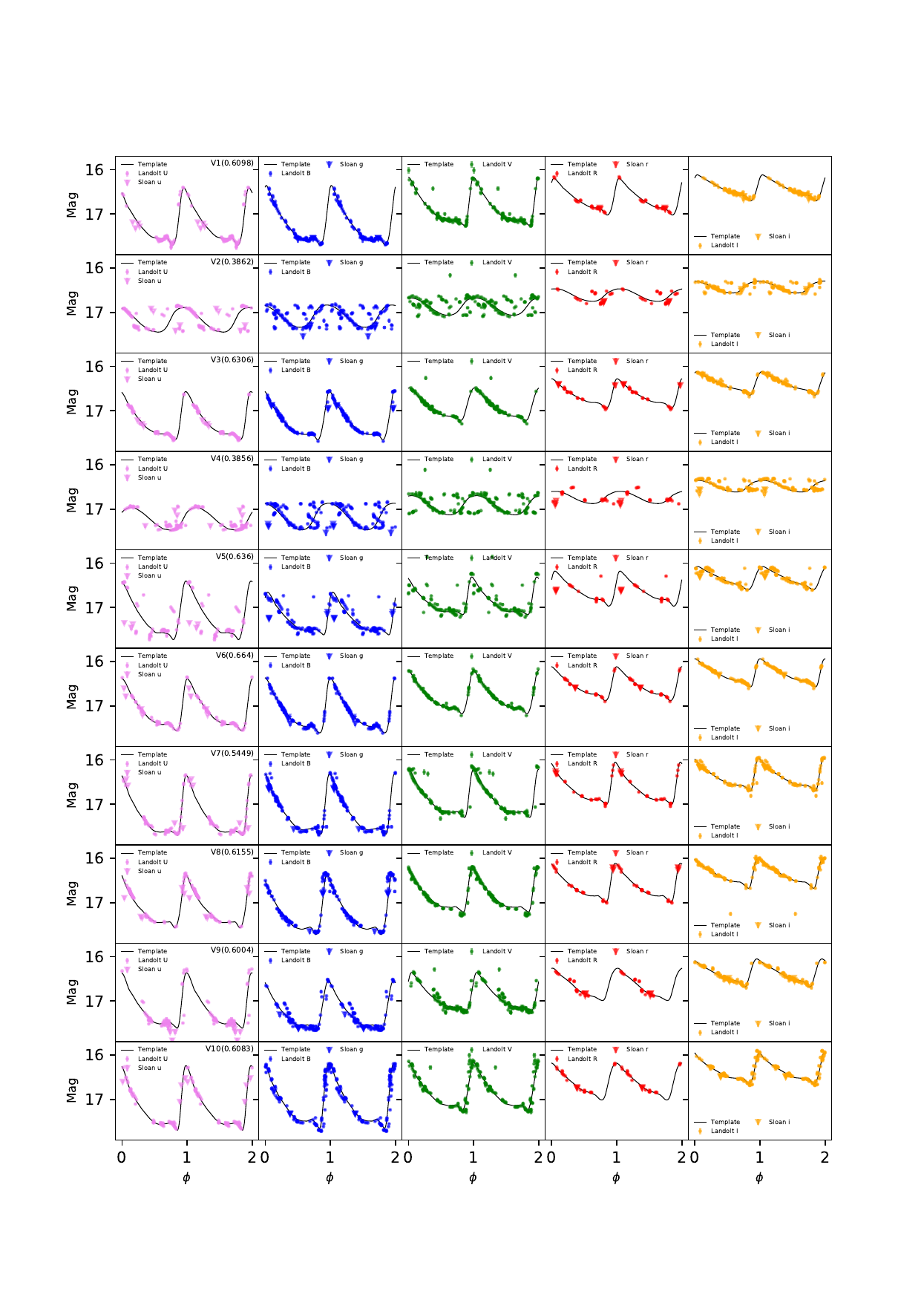}
    \caption{Template fits superimposed on the light curves for V1 to V10 in NGC 5024. Each row corresponds to a variable, with five columns representing filters in the order U, B, V, R, and I. Periods are mentioned in parentheses in the first column of each row.}
    \label{img:temp1to10}
\end{figure*}

\begin{figure*}    
    \includegraphics[width=0.9\linewidth]{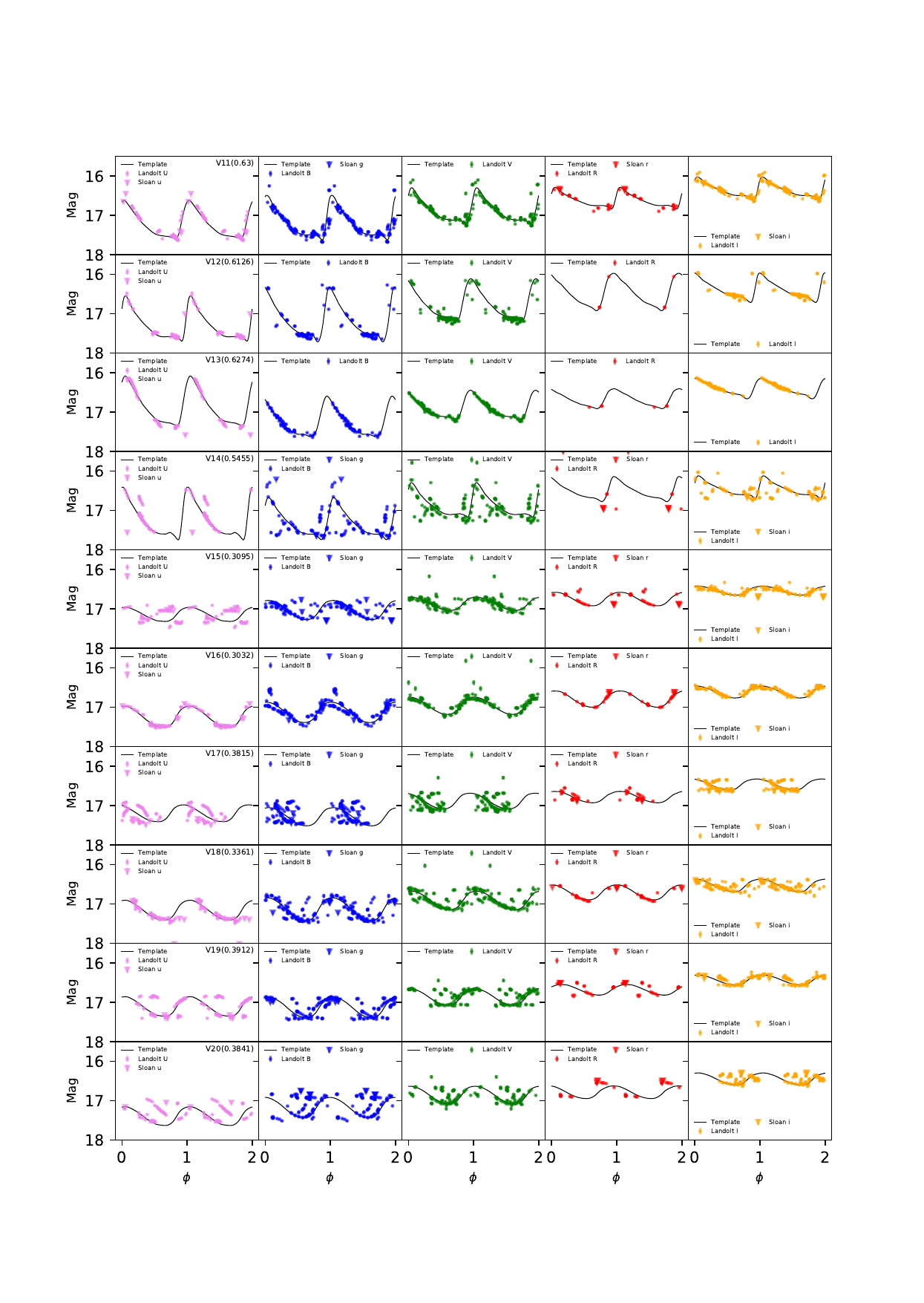}
    \caption{Same as Figure~\ref{img:temp1to10}, but for V11 to V20.}
    \label{img:tempv11to20}
\end{figure*}

\begin{figure*}
    \includegraphics[width=0.9\linewidth]{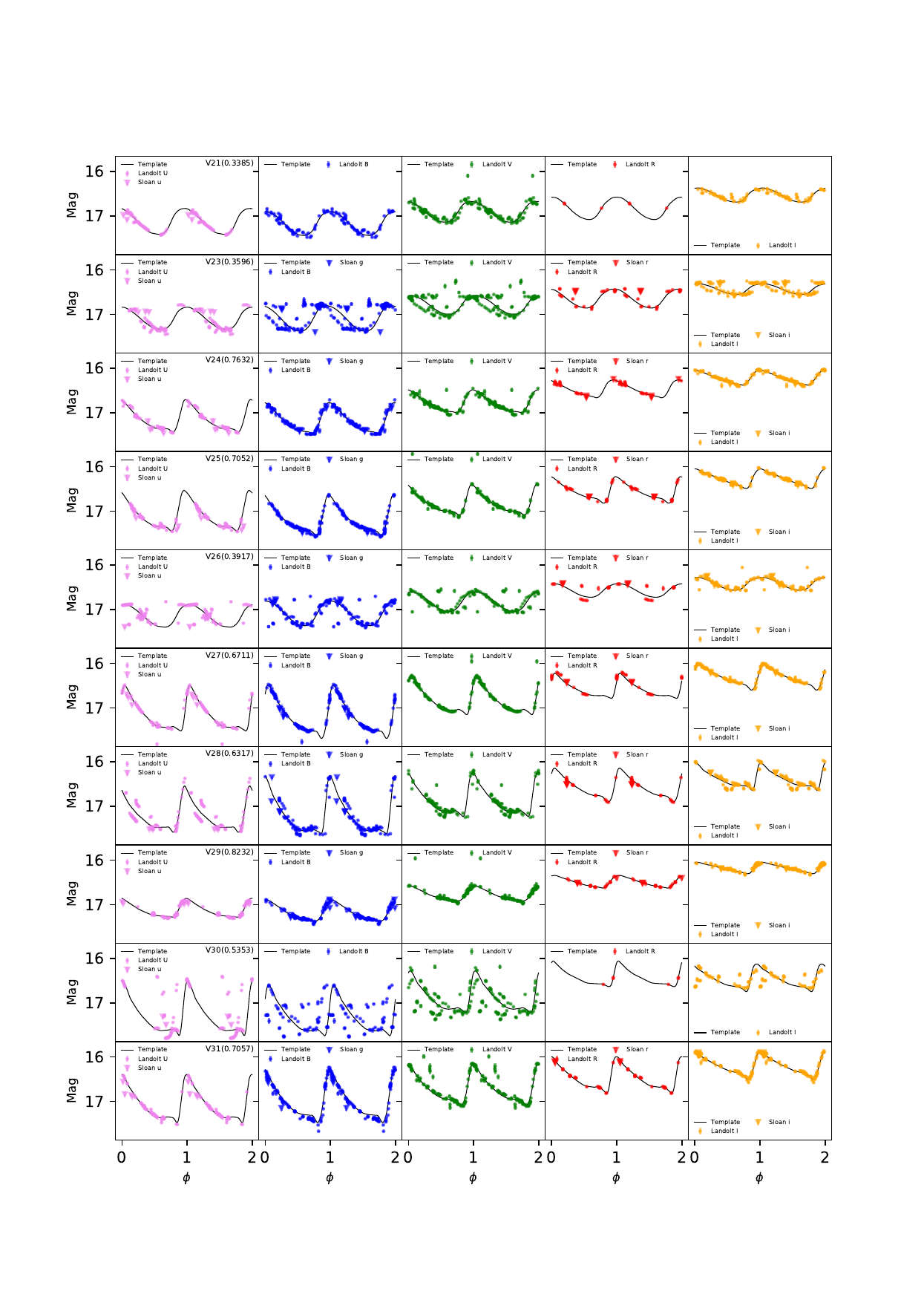}
    \caption{Same as Figure~\ref{img:temp1to10}, but for V21 to V31 (excluding V22).}
    \label{img:temp21to31}
\end{figure*}

\begin{figure*}    
    \includegraphics[width=0.9\linewidth]{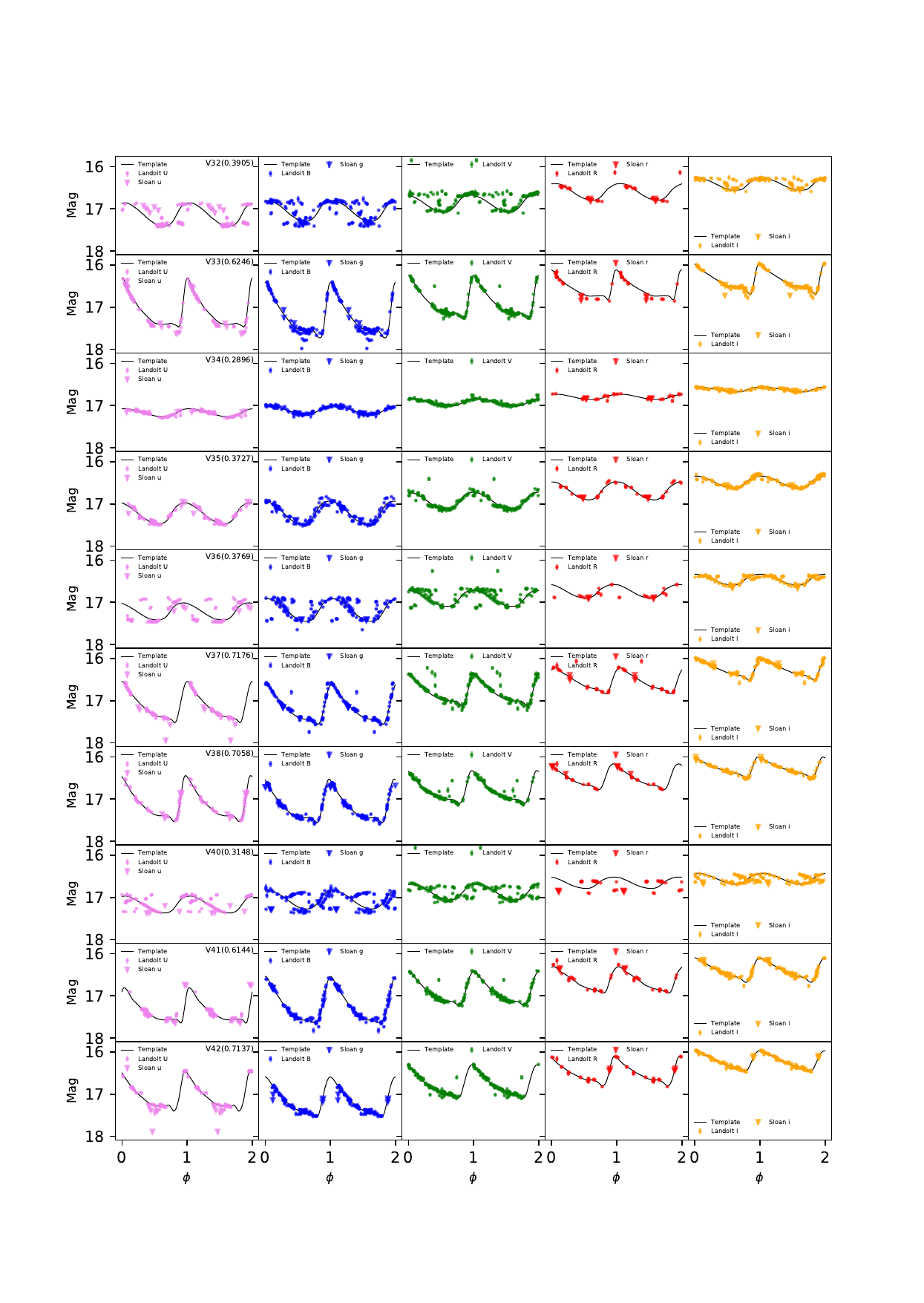}
    \caption{Same as Figure~\ref{img:temp1to10}, but for V32 to V42 (excluding V39).}
    \label{img:tempv32to42}
\end{figure*}

\begin{figure*}
    \includegraphics[width=0.9\linewidth]{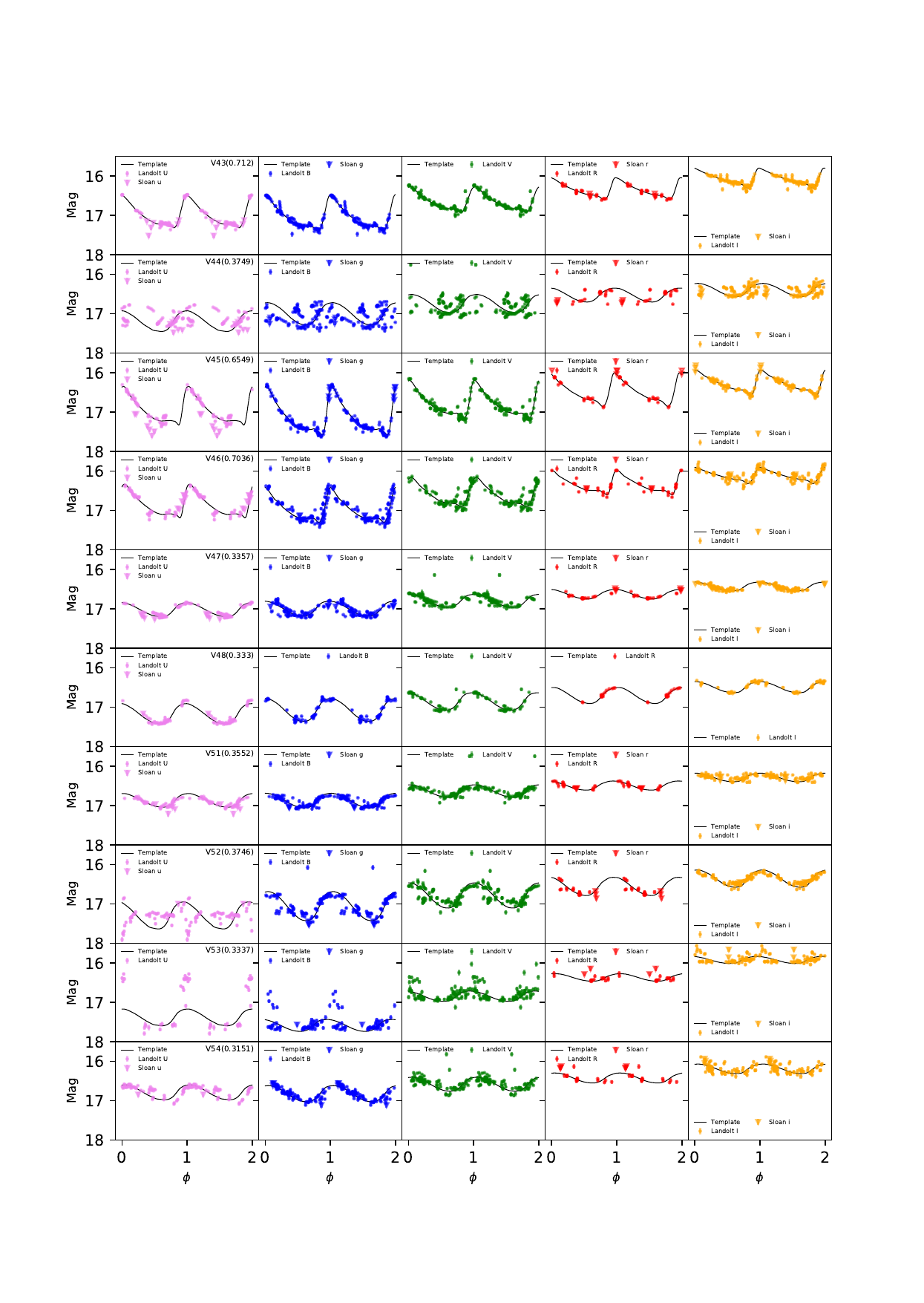}
    \caption{Same as Figure~\ref{img:temp1to10}, but for V43 to V54 (excluding V49 and V50).}
    \label{img:tempv43to54}
\end{figure*}

\begin{figure*}
    \includegraphics[width=0.9\linewidth]{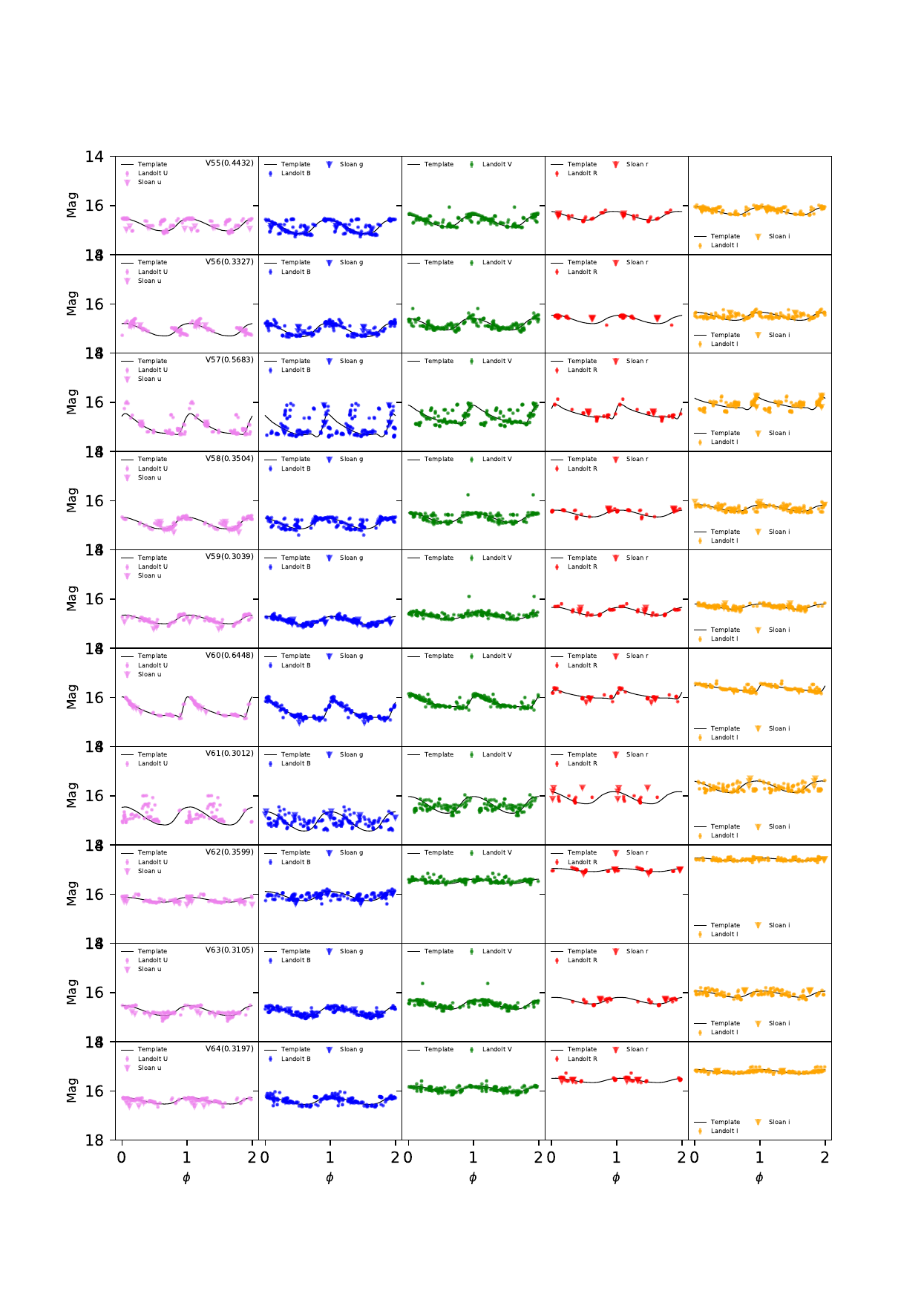}
    \caption{Same as Figure~\ref{img:temp1to10}, but for V55 to V64.}
    \label{img:tempv55to64}
\end{figure*}

\begin{figure*}
    \includegraphics[width=0.9\linewidth]{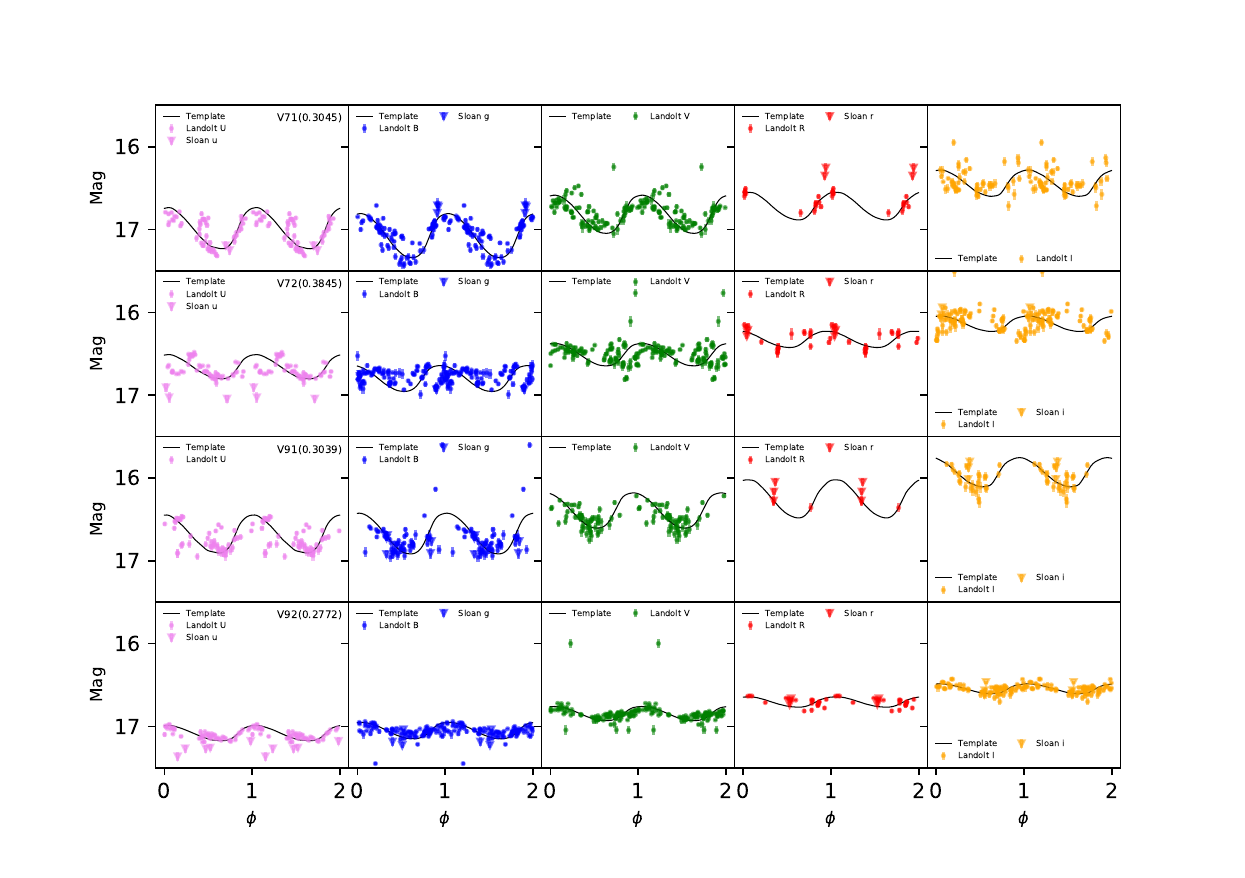}
    \caption{Same as Figure~\ref{img:temp1to10}, but for V71, V72, V91, and V92.}
    \label{img:tempv71to92}
\end{figure*}

The following figures provide supplementary information to support the main analysis presented in this paper. Figures~\ref{img:temp1to10} to \ref{img:tempv71to92} display the \textit{template-fitted light curves} for all RR Lyrae stars in our sample. These figures show the superimposed template fits on the observed data in different filters (U, B, V, R, and I) and are grouped by variable star number for clarity.

\begin{figure*}
    \centering
    \includegraphics[width=\linewidth]{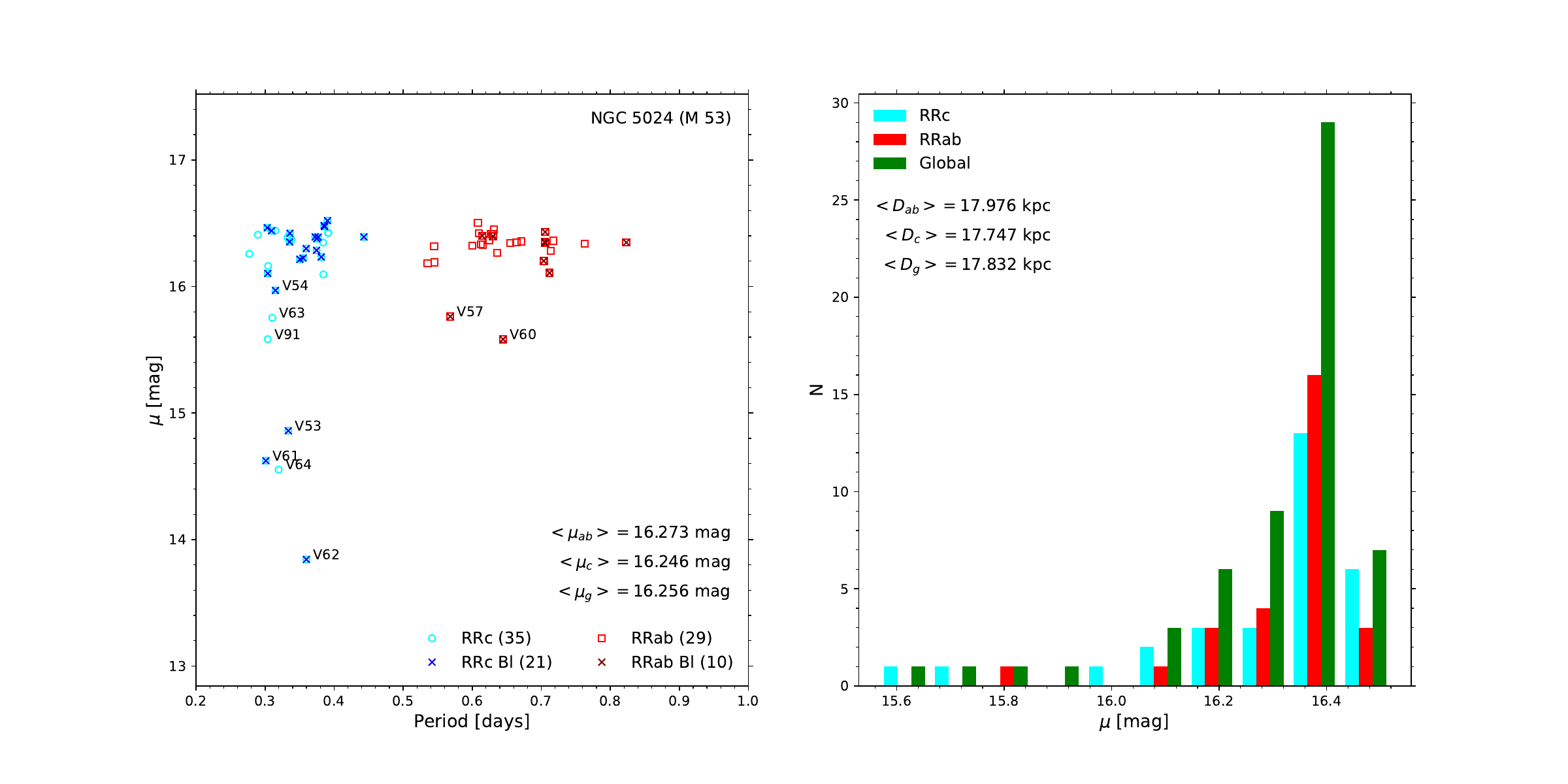}
    \caption{Distance modulus calculated using the (V, B-I) PWZ relation. The left panel shows RRc (cyan circles) and RRab (red squares), while the right panel presents a histogram of distance moduli for RRc, RRab, and Global PWZ relations. The weighted mean values of $\mu$ and the corresponding distances are indicated.}
    \label{img:muvbi}
\end{figure*}

\begin{figure*}
    \centering
    \includegraphics[width=\linewidth]{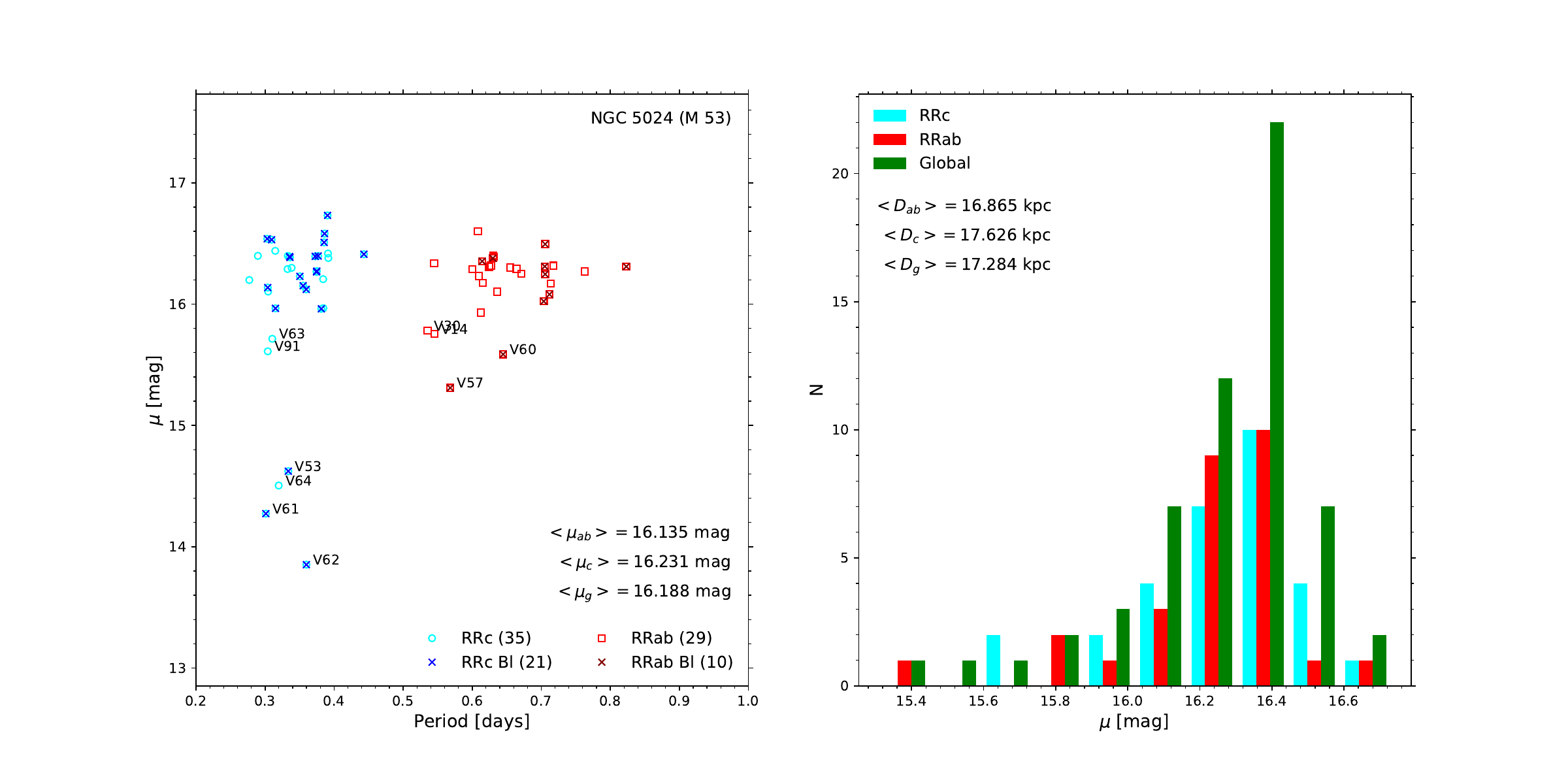}
    \caption{Same as Figure~\ref{img:muvbi}, but for the (V, B-V) PWZ relation.}
    \label{img:muvbv}
\end{figure*}

\begin{figure*}
    \centering
    \includegraphics[width=\linewidth]{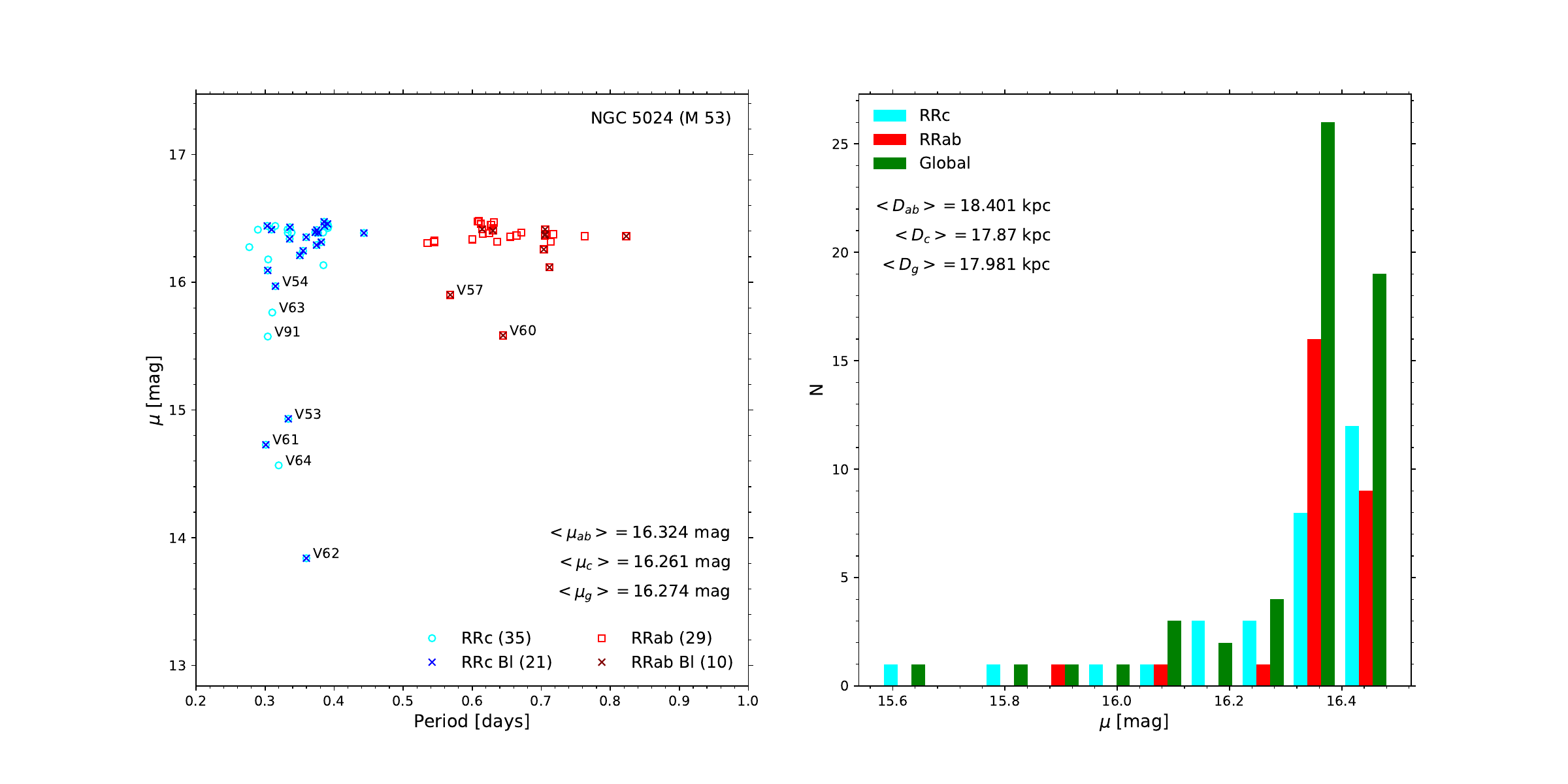}
    \caption{Same as Figure~\ref{img:muvbi}, but for the (I, B-I) PWZ relation.}
    \label{img:muibi}
\end{figure*}

\begin{figure*}
    \centering
    \includegraphics[width=\linewidth]{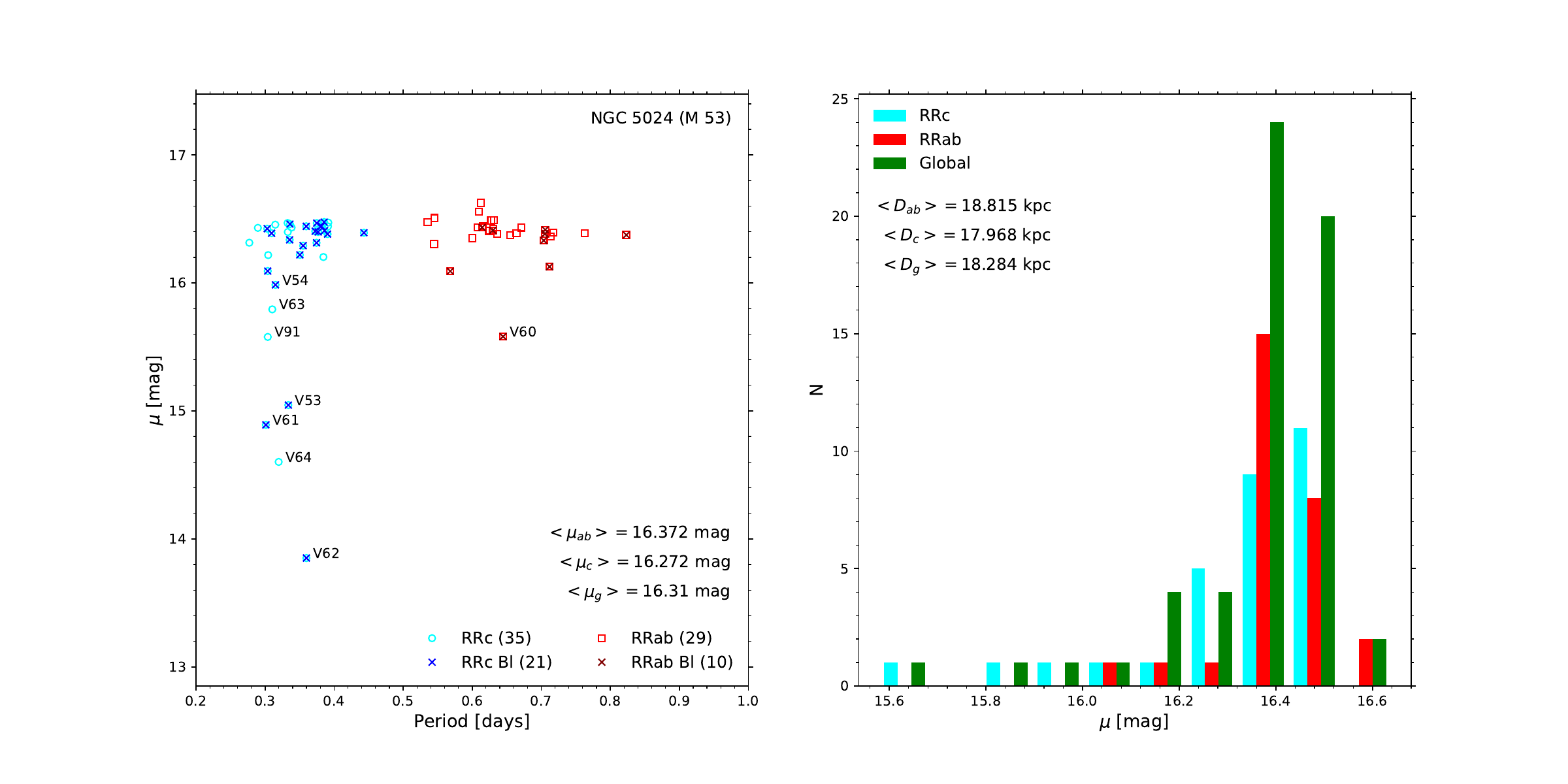}
    \caption{Same as Figure~\ref{img:muvbi}, but for the (I, V-I) PWZ relation.}
    \label{img:muivi}
\end{figure*}

\begin{figure*}
    \centering
    \includegraphics[width=\linewidth]{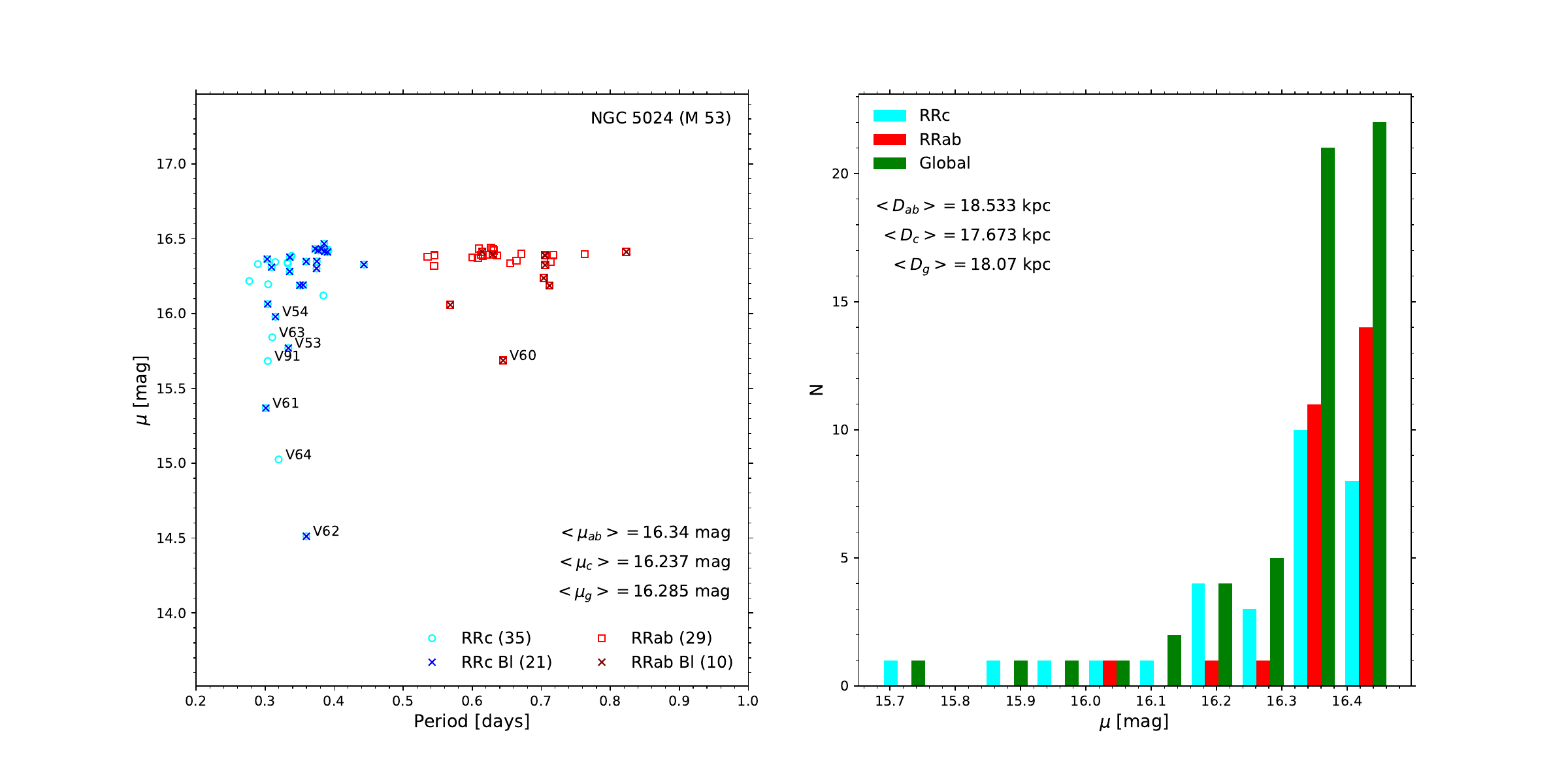}
    \caption{Same as Figure~\ref{img:muvbi}, but for the I-band PLZ relation.}
    \label{img:mui}
\end{figure*}

\begin{figure*}
    \centering
    \includegraphics[width=\linewidth]{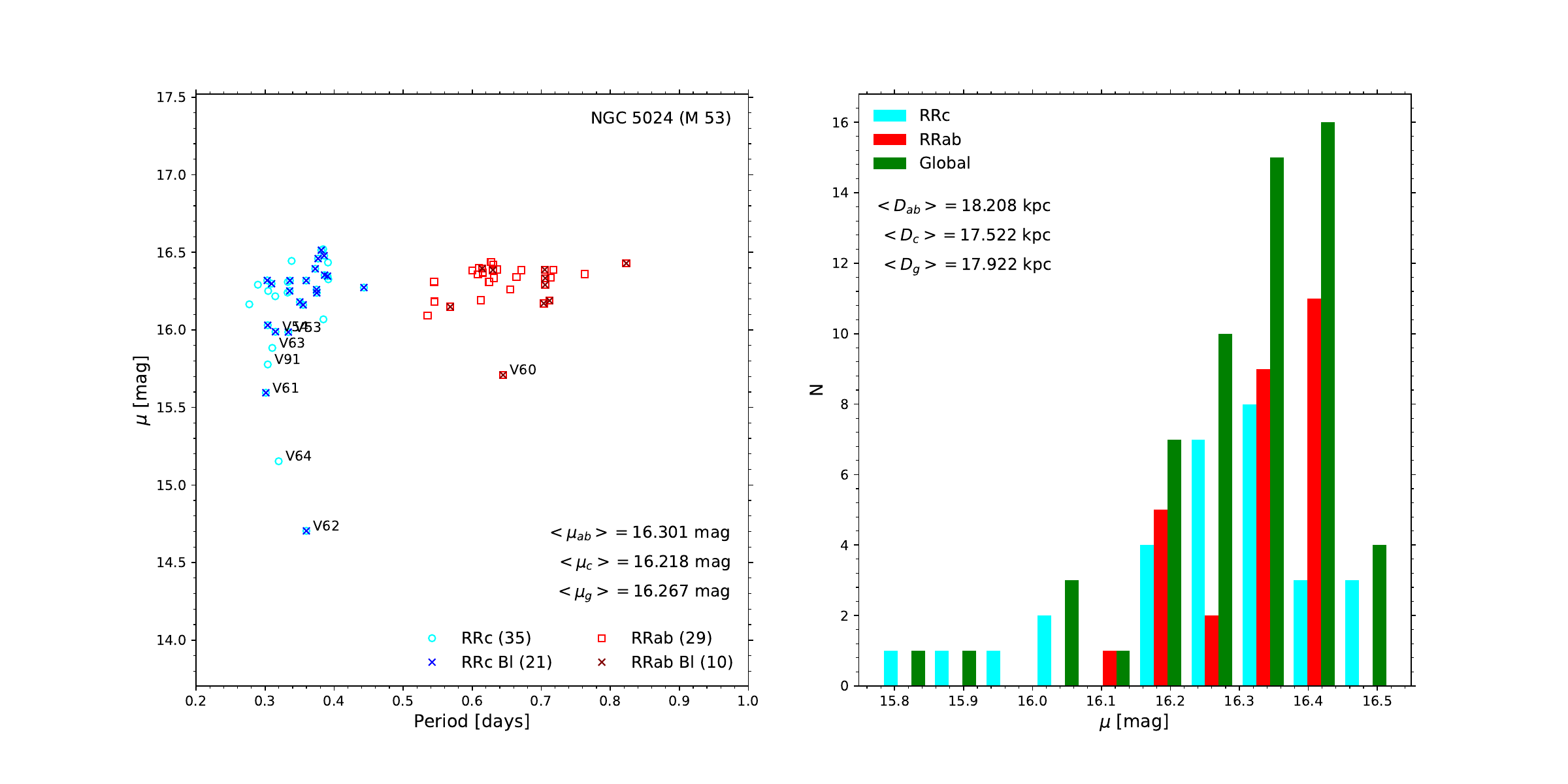}
    \caption{Same as Figure~\ref{img:muvbi}, but for the R-band PLZ relation.}
    \label{img:mur}
\end{figure*}

Figures~\ref{img:muvbi} through \ref{img:mur} show the \textit{distance modulus plots} derived using various \textit{Period-Wesenheit-Metallicity (PWZ)} and \textit{Period-Luminosity-Metallicity (PLZ)} relations. Each plot presents the calculated distance moduli for RRc and RRab stars, along with a histogram showing the distribution and weighted mean values.

\begin{figure*}
    \centering
    \includegraphics[width=\linewidth]{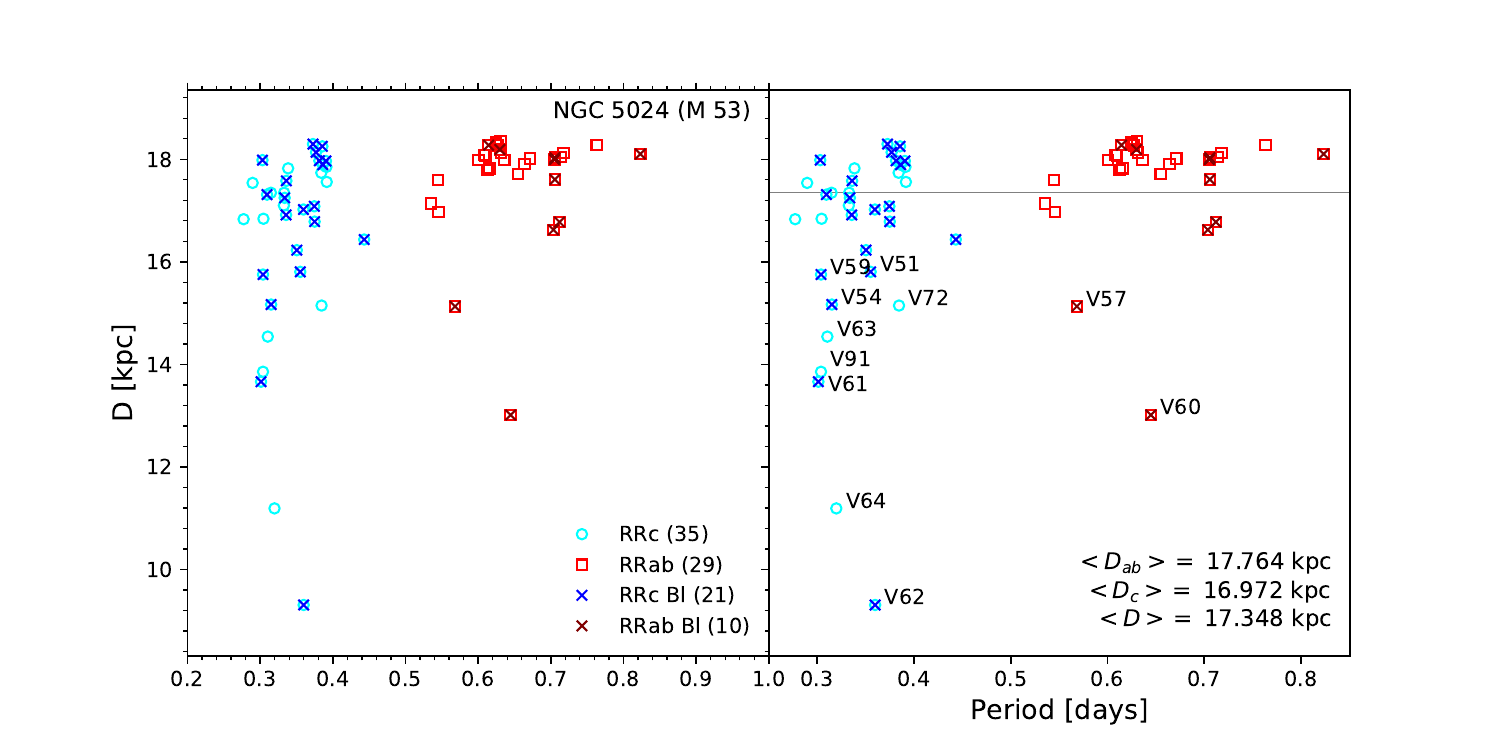}
    \caption{Distances obtained for RRc (cyan circles) and RRab (red squares) using Fourier-calculated absolute visual magnitudes. The weighted mean distances are indicated in the right panel. The grey line marks the overall mean distance.}
    \label{img:fourdist1}
\end{figure*}

Finally, Figure~\ref{img:fourdist1} shows the \textit{distances of RR Lyrae stars as a function of their period}, derived using mean absolute visual magnitudes calculated after Fourier fitting. The weighted mean distance for both RRab and RRc stars is highlighted to provide a visual summary of the results.

% Distance Modulus Plots

% Distance vs Period Plot

\section{Additional Tables}
Additional tables are provided here to supplement the main discussion. Table~\ref{tab:parameter10} and Table~\ref{tab:parameter11} list the photometric parameters for all the RR Lyrae candidates in NGC 5024, including their periods, types, mean magnitudes, and amplitudes in various bands.

\onecolumn
\begin{longtable}{cclllllcccccccc}
\captionsetup{justification=raggedright,singlelinecheck=false}
\caption{The table lists down the calculated periods, RR types, intensity averaged mean magnitudes, amplitudes, and the errors in the U, B, and V bands for all RRL candidates in NGC 5024.}
\label{tab:parameter10}
\\
\hline
Variable & Period & Type & \multicolumn{4}{c}{U-Band} & \multicolumn{4}{c}{B-Band} & \multicolumn{4}{c}{V-Band} \\
\hline
\endfirsthead

\multicolumn{15}{l}%
{{Table \thetable\ continued from previous page}} \\
\hline
Variable & Period & Type & \multicolumn{4}{c}{U-Band} & \multicolumn{4}{c}{B-Band} & \multicolumn{4}{c}{V-Band} \\
\hline
\endhead

\hline
\endfoot

\endlastfoot

& {[}Days] & & $U$ & $U_{err}$ & $A_U$ & $A_{U_{err}}$ & $B$ & $B_{err}$ & $A_B$ & $A_{B_{err}}$ & $V$ & $V_{err}$ & $A_V$ & $A_{V_{err}}$ \\
\hline
& & & \multicolumn{12}{c}{{[}mag]} \\
\hline
V1 & 0.60982 & RRab & 17.151 & 0.052 & 1.3 & 0.09 & 17.148 & 0.037 & 1.35 & 0.04 & 16.821 & 0.032 & 1.09 & 0.09 \\
V2 & 0.38620 & RRc & 17.159 & 0.029 & 0.55 & 0.07 & 17.069 & 0.019 & 0.5 & 0.1 & 16.86 & 0.016 & 0.39 & 0.06 \\
V3 & 0.63060 & RRab & 17.197 & 0.041 & 1.08 & 0.04 & 17.199 & 0.031 & 1.1 & 0.09 & 16.89 & 0.02 & 0.73 & 0.05 \\
V4 & 0.38558 & RRc & 17.211 & 0.027 & 0.51 & 0.09 & 17.144 & 0.021 & 0.6 & 0.09 & 16.901 & 0.019 & 0.43 & 0.08 \\
V5 & 0.63604 & RRab & 17.126 & 0.059 & 1.32 & 0.06 & 17.232 & 0.032 & 0.96 & 0.09 & 16.845 & 0.033 & 0.85 & 0.05 \\
V6 & 0.66402 & RRab & 17.065 & 0.038 & 1.16 & 0.06 & 17.08 & 0.032 & 1.24 & 0.05 & 16.76 & 0.03 & 0.97 & 0.07 \\
V7 & 0.54486 & RRab & 17.186 & 0.055 & 1.35 & 0.1 & 17.124 & 0.043 & 1.36 & 0.12 & 16.857 & 0.04 & 1.08 & 0.08 \\
V8 & 0.61552 & RRab & 17.09 & 0.048 & 1.2 & 0.08 & 17.159 & 0.044 & 1.36 & 0.13 & 16.814 & 0.036 & 0.99 & 0.15 \\
V9 & 0.60036 & RRab & 17.113 & 0.047 & 1.24 & 0.09 & 17.193 & 0.029 & 1.13 & 0.08 & 16.873 & 0.026 & 0.91 & 0.08 \\
V10 & 0.60826 & RRab & 17.09 & 0.057 & 1.46 & 0.1 & 17.012 & 0.049 & 1.46 & 0.19 & 16.803 & 0.038 & 1.1 & 0.12 \\
V11 & 0.62996 & RRab & 17.181 & 0.042 & 0.97 & 0.12 & 17.145 & 0.03 & 1.13 & 0.09 & 16.844 & 0.027 & 0.93 & 0.09 \\
V12 & 0.61259 & RRab & 17.24 & 0.051 & 1.15 & 0.06 & 17.084 & 0.05 & 1.4 & 0.04 & 16.701 & 0.032 & 1.08 & 0.04 \\
V13 & 0.62744 & RRab & 16.801 & 0.084 & 1.25 & 0.07 & 17.193 & 0.031 & 1.03 & 0.04 & 16.868 & 0.021 & 0.75 & 0.07 \\
V14 & 0.54549 & RRab & 17.163 & 0.049 & 1.34 & 0.04 & 17.269 & 0.046 & 1.07 & 0.08 & 16.834 & 0.032 & 0.88 & 0.12 \\
V15 & 0.30947 & RRc & 17.137 & 0.023 & 0.36 & 0.07 & 17.014 & 0.015 & 0.49 & 0.1 & 16.865 & 0.016 & 0.34 & 0.06 \\
V16 & 0.30317 & RRc & 17.232 & 0.028 & 0.54 & 0.05 & 17.113 & 0.021 & 0.52 & 0.16 & 16.948 & 0.021 & 0.44 & 0.11 \\
V17 & 0.38148 & RRc & 17.19 & 0.023 & 0.43 & 0.04 & 17.269 & 0.021 & 0.46 & 0.05 & 16.873 & 0.018 & 0.4 & 0.08 \\
V18 & 0.33606 & RRc & 17.092 & 0.026 & 0.6 & 0.07 & 17.087 & 0.022 & 0.67 & 0.09 & 16.865 & 0.02 & 0.55 & 0.07 \\
V19 & 0.39117 & RRc & 17.098 & 0.028 & 0.5 & 0.07 & 17.108 & 0.021 & 0.53 & 0.09 & 16.848 & 0.017 & 0.42 & 0.06 \\
V20 & 0.38414 & RRc & 17.384 & 0.025 & 0.48 & 0.08 & 17.154 & 0.025 & 0.51 & 0.09 & 16.84 & 0.019 & 0.44 & 0.12 \\
V21 & 0.33851 & RRc & 17.111 & 0.025 & 0.58 & 0.07 & 17.152 & 0.025 & 0.53 & 0.04 & 16.893 & 0.02 & 0.46 & 0.09 \\
V23 & 0.35962 & RRc & 17.082 & 0.027 & 0.5 & 0.05 & 17.072 & 0.024 & 0.57 & 0.17 & 16.775 & 0.02 & 0.42 & 0.09 \\
V24 & 0.76319 & RRab & 17.121 & 0.028 & 0.73 & 0.04 & 17.148 & 0.022 & 0.69 & 0.09 & 16.77 & 0.015 & 0.51 & 0.05 \\
V25 & 0.70515 & RRab & 17.057 & 0.025 & 0.91 & 0.07 & 17.147 & 0.023 & 0.92 & 0.1 & 16.8 & 0.022 & 0.69 & 0.11 \\
V26 & 0.39165 & RRc & 17.133 & 0.024 & 0.51 & 0.05 & 17.071 & 0.024 & 0.55 & 0.05 & 16.811 & 0.021 & 0.44 & 0.1 \\
V27 & 0.67107 & RRab & 17.128 & 0.045 & 1.02 & 0.07 & 17.153 & 0.036 & 1.21 & 0.06 & 16.803 & 0.031 & 0.85 & 0.08 \\
V28 & 0.63168 & RRab & 17.175 & 0.041 & 1.04 & 0.05 & 17.121 & 0.043 & 1.3 & 0.03 & 16.828 & 0.035 & 0.97 & 0.03 \\
V29 & 0.82325 & RRab & 17.116 & 0.019 & 0.42 & 0.09 & 17.161 & 0.015 & 0.47 & 0.04 & 16.769 & 0.015 & 0.34 & 0.09 \\
V30 & 0.53534 & RRab & 17.238 & 0.071 & 1.22 & 0.06 & 17.273 & 0.041 & 1.14 & 0.09 & 16.848 & 0.033 & 0.93 & 0.08 \\
V31 & 0.70567 & RRab & 17.02 & 0.039 & 1.09 & 0.08 & 16.943 & 0.039 & 1.2 & 0.1 & 16.685 & 0.032 & 0.94 & 0.11 \\
V32 & 0.39054 & RRc & 17.122 & 0.028 & 0.55 & 0.05 & 17.033 & 0.022 & 0.49 & 0.09 & 16.864 & 0.019 & 0.4 & 0.11 \\
V33 & 0.62459 & RRab & 17.039 & 0.054 & 1.16 & 0.07 & 17.163 & 0.045 & 1.31 & 0.06 & 16.844 & 0.032 & 1.02 & 0.06 \\
V34 & 0.28962 & RRc & 17.169 & 0.008 & 0.19 & 0.05 & 17.104 & 0.009 & 0.21 & 0.08 & 16.921 & 0.007 & 0.18 & 0.06 \\
V35 & 0.37267 & RRc & 17.214 & 0.021 & 0.49 & 0.06 & 17.191 & 0.02 & 0.55 & 0.07 & 16.92 & 0.017 & 0.42 & 0.11 \\
V36 & 0.37687 & RRc & 17.212 & 0.031 & 0.41 & 0.06 & 17.167 & 0.022 & 0.54 & 0.07 & 16.899 & 0.018 & 0.38 & 0.07 \\
V37 & 0.71762 & RRab & 17.085 & 0.041 & 0.98 & 0.04 & 17.134 & 0.029 & 0.95 & 0.07 & 16.787 & 0.025 & 0.78 & 0.11 \\
V38 & 0.70579 & RRab & 17.067 & 0.035 & 1.09 & 0.09 & 17.131 & 0.029 & 1.03 & 0.07 & 16.773 & 0.021 & 0.78 & 0.05 \\
V40 & 0.31482 & RRc & 17.165 & 0.02 & 0.4 & 0.07 & 17.044 & 0.018 & 0.43 & 0.13 & 16.862 & 0.018 & 0.37 & 0.07 \\
V41 & 0.61445 & RRab & 17.323 & 0.023 & 0.84 & 0.08 & 17.175 & 0.033 & 1.1 & 0.11 & 16.868 & 0.022 & 0.8 & 0.06 \\
V42 & 0.71372 & RRab & 17.005 & 0.044 & 0.95 & 0.05 & 17.108 & 0.021 & 0.94 & 0.05 & 16.733 & 0.023 & 0.78 & 0.08 \\
V43 & 0.71201 & RRab & 16.96 & 0.039 & 0.83 & 0.1 & 16.998 & 0.027 & 0.92 & 0.13 & 16.629 & 0.02 & 0.65 & 0.19 \\
V44 & 0.37494 & RRc & 17.18 & 0.022 & 0.52 & 0.09 & 16.98 & 0.02 & 0.56 & 0.07 & 16.728 & 0.02 & 0.46 & 0.08 \\
V45 & 0.65494 & RRab & 16.914 & 0.045 & 0.98 & 0.05 & 17.061 & 0.039 & 1.24 & 0.09 & 16.752 & 0.027 & 0.95 & 0.15 \\
V46 & 0.70364 & RRab & 16.834 & 0.033 & 0.86 & 0.15 & 16.942 & 0.029 & 0.89 & 0.16 & 16.576 & 0.025 & 0.74 & 0.18 \\
V47 & 0.33569 & RRc & 17.017 & 0.018 & 0.36 & 0.06 & 16.986 & 0.014 & 0.38 & 0.08 & 16.789 & 0.014 & 0.33 & 0.07 \\
V48 & 0.33297 & RRc & 17.157 & 0.03 & 0.51 & 0.1 & 17.048 & 0.033 & 0.58 & 0.05 & 16.841 & 0.027 & 0.44 & 0.06 \\
V51 & 0.35521 & RRc & 16.863 & 0.013 & 0.35 & 0.11 & 16.862 & 0.011 & 0.37 & 0.11 & 16.624 & 0.013 & 0.32 & 0.14 \\
V52 & 0.37458 & RRc & 17.272 & 0.029 & 0.69 & 0.2 & 17.02 & 0.027 & 0.74 & 0.18 & 16.757 & 0.02 & 0.64 & 0.17 \\
V53 & 0.33372 & RRc & 17.376 & 0.118 & 0.42 & 0.12 & 17.579 & 0.028 & 0.3 & 0.18 & 16.837 & 0.019 & 0.26 & 0.27 \\
V54 & 0.31511 & RRc & 16.782 & 0.017 & 0.38 & 0.11 & 16.808 & 0.014 & 0.41 & 0.12 & 16.571 & 0.016 & 0.35 & 0.23 \\
V55 & 0.44322 & RRc & 16.781 & 0.028 & 0.5 & 0.12 & 16.851 & 0.023 & 0.56 & 0.12 & 16.616 & 0.019 & 0.46 & 0.09 \\
V56 & 0.33265 & RRc & 17.039 & 0.025 & 0.51 & 0.07 & 17.038 & 0.022 & 0.55 & 0.14 & 16.808 & 0.019 & 0.47 & 0.13 \\
V57 & 0.5683 & RRab & 16.964 & 0.046 & 0.86 & 0.15 & 17.038 & 0.044 & 0.94 & 0.18 & 16.542 & 0.029 & 0.77 & 0.2 \\
V58 & 0.35045 & RRc & 16.91 & 0.026 & 0.47 & 0.13 & 16.911 & 0.018 & 0.44 & 0.17 & 16.682 & 0.017 & 0.38 & 0.13 \\
V59 & 0.30394 & RRc & 16.827 & 0.02 & 0.35 & 0.1 & 16.869 & 0.012 & 0.34 & 0.21 & 16.667 & 0.012 & 0.3 & 0.18 \\
V60 & 0.64476 & RRab & 16.494 & 0.035 & 0.84 & 0.08 & 16.566 & 0.027 & 0.82 & 0.15 & 16.209 & 0.019 & 0.57 & 0.14 \\
V61 & 0.30119 & RRc & 16.8 & 0.044 & 0.73 & 0.18 & 17.002 & 0.023 & 0.78 & 0.2 & 16.34 & 0.02 & 0.68 & 0.22 \\
V62 & 0.35986 & RRc & 16.224 & 0.014 & 0.2 & 0.1 & 16.067 & 0.011 & 0.37 & 0.13 & 15.477 & 0.008 & 0.18 & 0.16 \\
V63 & 0.31047 & RRc & 16.728 & 0.022 & 0.4 & 0.15 & 16.765 & 0.015 & 0.43 & 0.24 & 16.483 & 0.014 & 0.37 & 0.17 \\
V64 & 0.31975 & RRc & 16.402 & 0.012 & 0.26 & 0.08 & 16.396 & 0.013 & 0.29 & 0.1 & 15.912 & 0.01 & 0.24 & 0.13 \\
V71 & 0.30451 & RRc & 16.977 & 0.022 & 0.5 & 0.11 & 17.056 & 0.021 & 0.53 & 0.12 & 16.803 & 0.015 & 0.46 & 0.13 \\
V72 & 0.38453 & RRc & 16.657 & 0.016 & 0.29 & 0.13 & 16.792 & 0.008 & 0.31 & 0.22 & 16.506 & 0.013 & 0.27 & 0.15 \\
V91 & 0.30393 & RRc & 16.671 & 0.021 & 0.46 & 0.21 & 16.656 & 0.023 & 0.49 & 0.17 & 16.382 & 0.016 & 0.43 & 0.22 \\
V92 & 0.27722 & RRc & 17.082 & 0.01 & 0.18 & 0.09 & 17.048 & 0.007 & 0.2 & 0.08 & 16.843 & 0.011 & 0.17 & 0.09 \\
\hline
\end{longtable}

\begin{longtable}{cccccccccccccc}
\captionsetup{justification=raggedright,singlelinecheck=false}
\caption{The table lists down the calculated periods, RR types, intensity averaged mean magnitudes, amplitudes, and the errors in R and I bands, along with mean RMS scatter around the template, flag, and Blazhko classification, for all RRL candidates in NGC 5024.}
\label{tab:parameter11}
\\
\hline
Variable & Period & Type & \multicolumn{4}{c}{R-Band} & \multicolumn{4}{c}{I-Band} & \multicolumn{3}{c}{Template Parameters} \\
\hline
& & & R & $R_{err}$ & $A_R$ & $A_{R_{err}}$ & I & $I_{err}$ & $A_I$ & $A_{I_{err}}$ & RMS & Flag & Blazhko \\
\hline
& {[}Days] & & \multicolumn{8}{c}{{[}mag]} & {[}mag] & & \\
\hline
\endfirsthead

\multicolumn{14}{l}%
{{\bfseries Table \thetable\ continued from previous page}} \\
\hline
Variable & Period & Type & \multicolumn{4}{c}{R-Band} & \multicolumn{4}{c}{I-Band} & \multicolumn{3}{c}{Template Parameters} \\
\hline
& & & R & $R_{err}$ & $A_R$ & $A_{R_{err}}$ & I & $I_{err}$ & $A_I$ & $A_{I_{err}}$ & RMS & Flag & Blazhko \\
\hline
& {[}Days] & & \multicolumn{8}{c}{{[}mag]} & {[}mag] & & \\
\hline
\endhead

\hline
\endfoot

\endlastfoot

V1 & 0.60982 & RRab & 16.64 & 0.04 & 0.83 & 0.02 & 16.409 & 0.011 & 0.59 & 0.03 & 0.048 & A & - \\
V2 & 0.38620 & RRc & 16.605 & 0.024 & 0.28 & 0.03 & 16.424 & 0.01 & 0.27 & 0.07 & 0.159 & B & Bl \\
V3 & 0.63060 & RRab & 16.642 & 0.036 & 0.67 & 0.03 & 16.38 & 0.016 & 0.53 & 0.07 & 0.040 & A & - \\
V4 & 0.38558 & RRc & 16.731 & 0.024 & 0.28 & 0.03 & 16.477 & 0.011 & 0.26 & 0.06 & 0.166 & B & Bl \\
V5 & 0.63604 & RRab & 16.605 & 0.039 & 0.79 & 0.03 & 16.332 & 0.019 & 0.53 & 0.06 & 0.191 & B & - \\
V6 & 0.66402 & RRab & 16.532 & 0.039 & 0.77 & 0.02 & 16.266 & 0.019 & 0.63 & 0.07 & 0.041 & A & - \\
V7 & 0.54486 & RRab & 16.62 & 0.06 & 0.95 & 0.06 & 16.375 & 0.026 & 0.74 & 0.11 & 0.0921 & A & - \\
V8 & 0.61552 & RRab & 16.605 & 0.057 & 0.88 & 0.05 & 16.354 & 0.022 & 0.66 & 0.1 & 0.085 & A & - \\
V9 & 0.60036 & RRab & 16.635 & 0.037 & 0.72 & 0.01 & 16.36 & 0.016 & 0.63 & 0.07 & 0.092 & A & - \\
V10 & 0.60826 & RRab & 16.601 & 0.049 & 0.83 & 0.02 & 16.349 & 0.024 & 0.7 & 0.1 & 0.094 & A & - \\
V11 & 0.62996 & RRab & 16.608 & 0.036 & 0.55 & 0.04 & 16.347 & 0.019 & 0.58 & 0.09 & 0.097 & A & Bl \\
V12 & 0.61259 & RRab & 16.431 & 0.26 & 0.95 & 0.01 & 16.365 & 0.026 & 0.75 & 0.06 & 0.104 & B & - \\
V13 & 0.62744 & RRab & 16.662 & 0.01 & 0.51 & 0.01 & 16.396 & 0.014 & 0.52 & 0.03 & 0.065 & A & - \\
V14 & 0.54549 & RRab & 16.491 & 0.201 & 0.64 & 0.04 & 16.445 & 0.025 & 0.59 & 0.06 & 0.300 & C & - \\
V15 & 0.30947 & RRc & 16.741 & 0.028 & 0.34 & 0.01 & 16.531 & 0.009 & 0.23 & 0.04 & 0.141 & B & Bl \\
V16 & 0.30317 & RRc & 16.78 & 0.03 & 0.41 & 0.01 & 16.605 & 0.01 & 0.3 & 0.05 & 0.068 & A & Bl \\
V17 & 0.38148 & RRc & 16.776 & 0.023 & 0.29 & 0.02 & 16.453 & 0.012 & 0.28 & 0.08 & 0.154 & C & Bl \\
V18 & 0.33606 & RRc & 16.69 & 0.027 & 0.46 & 0.03 & 16.519 & 0.01 & 0.32 & 0.06 & 0.119 & B & Bl \\
V19 & 0.39117 & RRc & 16.675 & 0.025 & 0.27 & 0.02 & 16.425 & 0.011 & 0.25 & 0.08 & 0.142 & B & - \\
V20 & 0.38414 & RRc & 16.776 & 0.039 & 0.32 & 0.01 & 16.438 & 0.012 & 0.31 & 0.07 & 0.225 & C & - \\
V21 & 0.33851 & RRc & 16.81 & 0.035 & 0.5 & 0.01 & 16.521 & 0.012 & 0.32 & 0.07 & 0.048 & A & - \\
V23 & 0.35962 & RRc & 16.631 & 0.032 & 0.42 & 0.06 & 16.425 & 0.013 & 0.24 & 0.08 & 0.132 & B & Bl \\
V24 & 0.76319 & RRab & 16.466 & 0.026 & 0.4 & 0.06 & 16.211 & 0.013 & 0.35 & 0.08 & 0.040 & A & - \\
V25 & 0.70515 & RRab & 16.542 & 0.032 & 0.6 & 0.05 & 16.261 & 0.012 & 0.44 & 0.06 & 0.049 & A & Bl \\
V26 & 0.39165 & RRc & 16.565 & 0.032 & 0.3 & 0.02 & 16.415 & 0.013 & 0.29 & 0.06 & 0.125 & B & - \\
V27 & 0.67107 & RRab & 16.57 & 0.038 & 0.57 & 0.06 & 16.305 & 0.018 & 0.59 & 0.06 & 0.059 & A & - \\
V28 & 0.63168 & RRab & 16.555 & 0.099 & 0.77 & 0.07 & 16.37 & 0.024 & 0.62 & 0.06 & 0.154 & B & - \\
V29 & 0.82325 & RRab & 16.49 & 0.014 & 0.28 & 0.01 & 16.171 & 0.008 & 0.26 & 0.05 & 0.036 & A & Bl \\
V30 & 0.53534 & RRab & 16.412 & 0.049 & 0.58 & 0.02 & 16.448 & 0.021 & 0.62 & 0.06 & 0.250 & C & - \\
V31 & 0.70567 & RRab & 16.444 & 0.044 & 0.82 & 0.05 & 16.194 & 0.023 & 0.61 & 0.12 & 0.065 & A & Bl \\
V32 & 0.39054 & RRc & 16.588 & 0.035 & 0.39 & 0.05 & 16.41 & 0.01 & 0.29 & 0.1 & 0.131 & C & Bl \\
V33 & 0.62459 & RRab & 16.536 & 0.051 & 0.71 & 0.04 & 16.35 & 0.025 & 0.72 & 0.13 & 0.065 & A & - \\
V34 & 0.28962 & RRc & 16.793 & 0.011 & 0.13 & 0.01 & 16.615 & 0.004 & 0.12 & 0.08 & 0.028 & A & - \\
V35 & 0.37267 & RRc & 16.676 & 0.03 & 0.43 & 0.03 & 16.475 & 0.013 & 0.29 & 0.04 & 0.051 & A & Bl \\
V36 & 0.37687 & RRc & 16.732 & 0.029 & 0.32 & 0.03 & 16.455 & 0.011 & 0.26 & 0.07 & 0.147 & B & Bl \\
V37 & 0.71762 & RRab & 16.53 & 0.048 & 0.62 & 0.04 & 16.249 & 0.017 & 0.53 & 0.07 & 0.079 & A & - \\
V38 & 0.70579 & RRab & 16.487 & 0.037 & 0.62 & 0.03 & 16.258 & 0.017 & 0.51 & 0.08 & 0.040 & A & Bl \\
V40 & 0.31482 & RRc & 16.646 & 0.023 & 0.27 & 0.05 & 16.549 & 0.008 & 0.26 & 0.1 & 0.157 & C & - \\
V41 & 0.61445 & RRab & 16.632 & 0.043 & 0.62 & 0.04 & 16.382 & 0.015 & 0.58 & 0.08 & 0.053 & A & Bl \\
V42 & 0.71372 & RRab & 16.483 & 0.042 & 0.7 & 0.03 & 16.209 & 0.016 & 0.48 & 0.08 & 0.051 & A & - \\
V43 & 0.71201 & RRab & 16.337 & 0.025 & 0.55 & 0.04 & 16.05 & 0.012 & 0.51 & 0.22 & 0.052 & A & Bl \\
V44 & 0.37494 & RRc & 16.516 & 0.026 & 0.35 & 0.04 & 16.387 & 0.013 & 0.34 & 0.1 & 0.214 & C & Bl \\
V45 & 0.65494 & RRab & 16.459 & 0.062 & 0.85 & 0.06 & 16.261 & 0.017 & 0.63 & 0.1 & 0.067 & A & - \\
V46 & 0.70364 & RRab & 16.326 & 0.037 & 0.62 & 0.12 & 16.109 & 0.016 & 0.41 & 0.21 & 0.091 & B & Bl \\
V47 & 0.33569 & RRc & 16.624 & 0.016 & 0.24 & 0.05 & 16.424 & 0.008 & 0.23 & 0.08 & 0.061 & A & Bl \\
V48 & 0.33297 & RRc & 16.689 & 0.029 & 0.41 & 0.01 & 16.484 & 0.021 & 0.28 & 0.08 & 0.044 & A & - \\
V51 & 0.35521 & RRc & 16.485 & 0.013 & 0.23 & 0.03 & 16.28 & 0.006 & 0.22 & 0.1 & 0.060 & A & Bl \\
V52 & 0.37458 & RRc & 16.538 & 0.029 & 0.46 & 0.09 & 16.34 & 0.013 & 0.44 & 0.24 & 0.172 & C & Bl \\
V53 & 0.33372 & RRc & 16.364 & 0.019 & 0.19 & 0.14 & 15.919 & 0.011 & 0.18 & 0.16 & 0.214 & C & Bl \\
V54 & 0.31511 & RRc & 16.416 & 0.033 & 0.26 & 0.09 & 16.183 & 0.013 & 0.24 & 0.17 & 0.108 & B & Bl \\
V55 & 0.44322 & RRc & 16.405 & 0.026 & 0.35 & 0.07 & 16.205 & 0.012 & 0.32 & 0.09 & 0.137 & B & Bl \\
V56 & 0.33265 & RRc & 16.621 & 0.017 & 0.34 & 0.1 & 16.495 & 0.01 & 0.33 & 0.14 & 0.139 & C & - \\
V57 & 0.5683 & RRab & 16.433 & 0.047 & 0.6 & 0.12 & 16.084 & 0.021 & 0.51 & 0.2 & 0.267 & C & Bl \\
V58 & 0.35045 & RRc & 16.516 & 0.023 & 0.28 & 0.07 & 16.291 & 0.013 & 0.27 & 0.12 & 0.109 & C & Bl \\
V59 & 0.30394 & RRc & 16.489 & 0.028 & 0.32 & 0.1 & 16.302 & 0.009 & 0.21 & 0.16 & 0.078 & A & Bl \\
V60 & 0.64476 & RRab & 15.917 & 0.034 & 0.39 & 0.07 & 15.622 & 0.013 & 0.34 & 0.13 & 0.069 & C & Bl \\
V61 & 0.30119 & RRc & 16.062 & 0.051 & 0.49 & 0.11 & 15.616 & 0.016 & 0.47 & 0.2 & 0.293 & C & Bl \\
V62 & 0.35986 & RRc & 15.017 & 0.012 & 0.13 & 0.04 & 14.589 & 0.005 & 0.12 & 0.14 & 0.086 & C & Bl \\
V63 & 0.31047 & RRc & 16.325 & 0.023 & 0.27 & 0.1 & 16.059 & 0.013 & 0.26 & 0.16 & 0.098 & B & - \\
V64 & 0.31975 & RRc & 15.568 & 0.016 & 0.18 & 0.02 & 15.215 & 0.006 & 0.17 & 0.11 & 0.076 & B & - \\
V71 & 0.30451 & RRc & 16.709 & 0.036 & 0.33 & 0.1 & 16.432 & 0.019 & 0.32 & 0.18 & 0.130 & B & - \\
V72 & 0.38453 & RRc & 16.323 & 0.023 & 0.2 & 0.1 & 16.134 & 0.015 & 0.19 & 0.18 & 0.131 & C & - \\
V91 & 0.30393 & RRc & 16.237 & 0.054 & 0.46 & 0.08 & 15.921 & 0.023 & 0.35 & 0.2 & 0.129 & B & - \\
V92 & 0.27722 & RRc & 16.704 & 0.01 & 0.12 & 0.05 & 16.544 & 0.005 & 0.12 & 0.13 & 0.071 & A & - \\
\hline
\end{longtable}
\twocolumn

\bibliographystyle{elsarticle-harv} 
\bibliography{paper}

%% else use the following coding to input the bibitems directly in the
%% TeX file.

%%\begin{thebibliography}{00}

%% \bibitem[Author(year)]{label}
%% For example:

%% \bibitem[Aladro et al.(2015)]{Aladro15} Aladro, R., Martín, S., Riquelme, D., et al. 2015, \aas, 579, A101

%%\end{thebibliography}

\end{document}